\documentclass[showpacs,amsmath,notitlepage,amssymb,aps,showkeys,floatfix,
prd,a4paper,onecolumn,nofootinbib]{revtex4} 
\usepackage[utf8]{inputenc} %
\usepackage{verbatim}    
\usepackage[dvips]{graphicx}
\usepackage{dcolumn}
\usepackage{bm}
\usepackage{epsfig}
\usepackage{braket}
\usepackage{epstopdf}
\usepackage{amsfonts}
\usepackage{amssymb,amscd}
\usepackage{color} 
\usepackage[normalem]{ulem} 
\usepackage{longtable} 

\newcommand{\be}{\begin{equation}}
\newcommand{\ee}{\end{equation}}
\newcommand{\ben}{\begin{eqnarray}}
\newcommand{\een}{\end{eqnarray}}

\newcommand{\mat}[1]{\mbox{\boldmath{$#1$}}} 
\newcommand{\Mask}{\sqrt{1+s_k}}
\newcommand{\Masq}{\sqrt{1+s_q}}
\newcommand{\Menosk}{\sqrt{1-s_k}}
\newcommand{\Menosq}{\sqrt{1-s_q}}
\newcommand{\gk}{\boldsymbol{\sigma}\cdot\boldsymbol{\hat{k}}}
\newcommand{\gq}{\boldsymbol{\sigma}\cdot\boldsymbol{\hat{q}}}

\newcommand{\dok}{\text{ d$\Omega_k$}}
\newcommand{\doq}{\text{ d$\Omega_q$}}
\newcommand{\bk}{\boldsymbol{k}}
\newcommand{\bq}{\boldsymbol{q}}
\begin{document}

\title{Mixing and $m_q$ dependence of axial vector mesons in the Coulomb gauge QCD model}

\author{ Luciano M. Abreu~\footnote{luciano.abreu@ufba.br} and Aline~G.~Favero}   
\affiliation{Instituto de F\'isica, Universidade Federal da Bahia, 
Campus Universit\'ario de Ondina, Salvador, Bahia, 40170-115, Brazil}
%
\author{ Felipe J. Llanes-Estrada\footnote{fllanes@ucm.es} and Alejandro Garc\'{\i}a S\'anchez}
\affiliation{ Departamento de F\'isica Te\'orica and IPARCOS, Universidad Complutense, Madrid, 28040, Spain}
%

\begin{abstract}
We discuss pure $q\overline{q}$ axial--vector mesons in the Tamm-Dancoff approximation
of the Coulomb--gauge QCD model from NCSU. While recent studies have put emphasis in 
configuration mixing with open meson--meson channels, we here concentrate on the simpler
closed--channel problem and follow the $1^+$ mixing through a wide range of quark masses.
We also examine their radial excitations and discuss with them the concept of insensitivity to 
chiral symmetry breaking.
\end{abstract}
\maketitle

\section{Introduction}
\label{sec:Introduction}
\subsection{$q\bar{q}$ as a rough guide to the spectrum}

\begin{figure}
\begin{minipage}{0.45\textwidth}
\begin{center}
\includegraphics*[width=\textwidth]{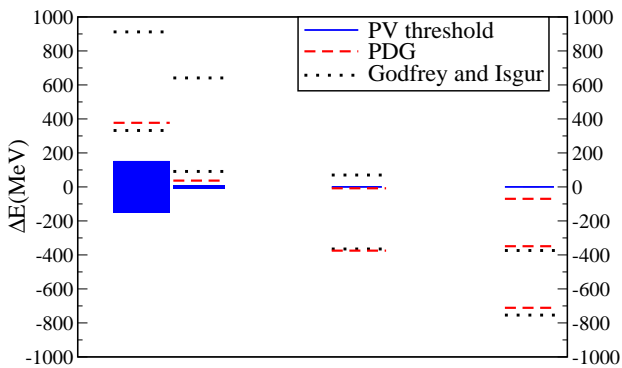}
\end{center}
\end{minipage}\ \ \
\begin{minipage}{0.45\textwidth}
\begin{center}
\caption{\label{fig:isgur} Spectrum of low--lying closed flavor axial--vector mesons. 
We compare the old predictions of the Godfrey and Isgur quark model~\cite{Godfrey:1985xj} with the current (central value) masses as listed in the Review of Particle Physics~\cite{Tanabashi:2018oca}. From left to right, the $q\bar{q}$ flavors are light--antilight, $s\bar{s}$, $c\bar{c}$, and $b\bar{b}$. 
In all cases the zero is normalized to the relevant vector--pseudoscalar decay threshold of the same quantum numbers.} 
\end{center}
\end{minipage}
\end{figure}

In this article we concentrate on axial--vector mesons in a $q\bar{q}$ field theory approach.
It is natural to question, at a time where exotic and hidden exotic mesons are widely discussed,
why is a discussion limited to quark--antiquark configurations even thinkable. 
Therefore, we plot in figure~\ref{fig:isgur} the two lowest traditional quark--model states for each quark flavor (dotted lines, from~\cite{Godfrey:1985xj}) against the experimental states~\cite{Tanabashi:2018oca}, and all shifted in mass so that the relevant $0^-1^-$ s--wave threshold is at $E=0$ (hence $\pi\rho$, $KK^*$, $DD^*$ and $BB^*$ all appear at the same height in the spectral Grotrian diagram: this removes the additive effect of the quark mass).
The figure shows several well--known features: that heavy quark states are more deeply bound and the one nearest decay threshold can be a radial excitation (the first one for charm, the second one for bottom); that the $c\bar{c}$ state, the renowned $X(3872)$ is a bit low as compared to the pure $q\bar{q}$ model prediction, and just at the decay threshold; and that, because the pions are so light and hence the threshold so low, excited $1^+$ states made of light quarks are broad and extremely difficult to reconstruct in experiment.

But most importantly, it shows that the quark model gets the basic picture right, roughly identifying where the different axial vector mesons should be. Of course, coupling to meson--meson channels can profoundly change the properties of any one particular state. But to study global properties of the spectrum, it is clear that the $q\bar{q}$ approach, even without that claim to precision in any particular state, is sensible.

Mesons are eigenstates of parity. In the quark model, a quark and an antiquark in the cm frame
have total orbital angular momentum equal to that of the relative particle, $L=l$, and the parity 
is  $P:=(-1)^{L+1}$. Positive parity is thus achieved with odd orbital angular momentum. 

Further, if the $q$ and $\bar{q}$ are of the same (opposite) flavor, then the meson is an eigenstate of charge conjugation, which is given by the total spin $S=s_q+s_{\bar{q}}$ as $C:=-(-1)^L(-1)^{S+1}=(-1)^{L+S}$.
Because $S$ can only take the values $S=0$ and $S=1$, positive charge conjugation implies 
$S=1$ (because $L$ is odd) and the triangle inequality forces $L=1$. These $J^{PC}=1^{++}$ mesons are then of necessity eigenstates of $L$ and $S$ with respective eigenvalues $1$ and $1$ (spin triplet). In the traditional spectroscopic notation, they are $^3P_1$. Similarly, with $P=+1$ but $C=-1$ we must have
$L=1$, $S=0$ or $^1P_1$. 

This pure $L$--$S$ basis therefore diagonalizes the infinitely heavy quarkonium and is a very good approximation for the $c\bar{c}$ and $b\bar{b}$ spectrum. For light mesons whose quantum numbers are compatible with a quark composition $s\bar{s}$ and $n\bar{n}$ (with $n=u,d$), there is no reason to expect that $q\bar{q}$ appropriately reflects the underlying meson structure given that the strong interactions can create an arbitrary number of light quarks and gluons. Nevertheless, the counting of states, their quantum numbers, and their approximate position in the spectrum follows the naive quark model counting: the only prominent exotic multiplet is Jaffe's inverted scalar nonet~\cite{Pelaez:2015qba}.

 This unreasonable agreement has led to the formulation of field--theory based quasiparticle approximations in which the bare quarks are dressed by $q\bar{q}$ pairs as in the BCS mechanism~\cite{Adler:1984ri}, or by gluons modeling the QCD Dyson--Schwinger equations~\cite{Fischer:2003rp}. The idea is that a quark mass--gap dominates most of the low--lying spectrum that therefore admits a description in terms of only a quark and an antiquark. This is consistent with chiral symmetry breaking and the Goldstone boson nature of the pion and kaon.

\subsection{Mixing of $1^+$ mesons}

For open--flavor mesons, in which the quark and antiquark have different flavor (unlike in quarkonium), charge conjugation is no more a symmetry. Therefore, even in simple quark models, there is no reason to expect that $S$ is a good quantum number. Here, the ${\bf L}$--${\bf S}$ states are still an optional basis, but since both $^3P_1$ and $^1P_1$ have equal quantum numbers $J^P=1^+$, they generally mix. 

There is an extreme case when either the quark (or the antiquark) is much heavier than its partner (or generically, when its mass accounts for much of the meson's). Then, Heavy Quark Symmetry applies and we know that the spin of the heavy quark is a good quantum number because it cannot be reversed. Then the correct way of building total $J$ is by coupling first the light partner spin and the angular momentum into its total ${\bf j}_q={\bf s}_q+{\bf l}$ and then couple this to the heavy quark's spin ${\bf J}={\bf s}_Q+{\bf j}_q$. The states can then be labeled as $(s_Q,j_q)_J$. In the heavy quark limit, $m_Q\gg m_q$, these are good quantum numbers. 

For the intermediate case where the masses are different, $m_f\neq m_{f'}$, neither set is made of good quantum numbers and we can speak (if only two states are considered) of a mixing angle $\theta_P$ referred to the ${\bf L}$--${\bf S}$ basis. 

\begin{figure}
\begin{minipage}{0.45\textwidth}
\begin{center}
\frame{\includegraphics*[width=\textwidth]{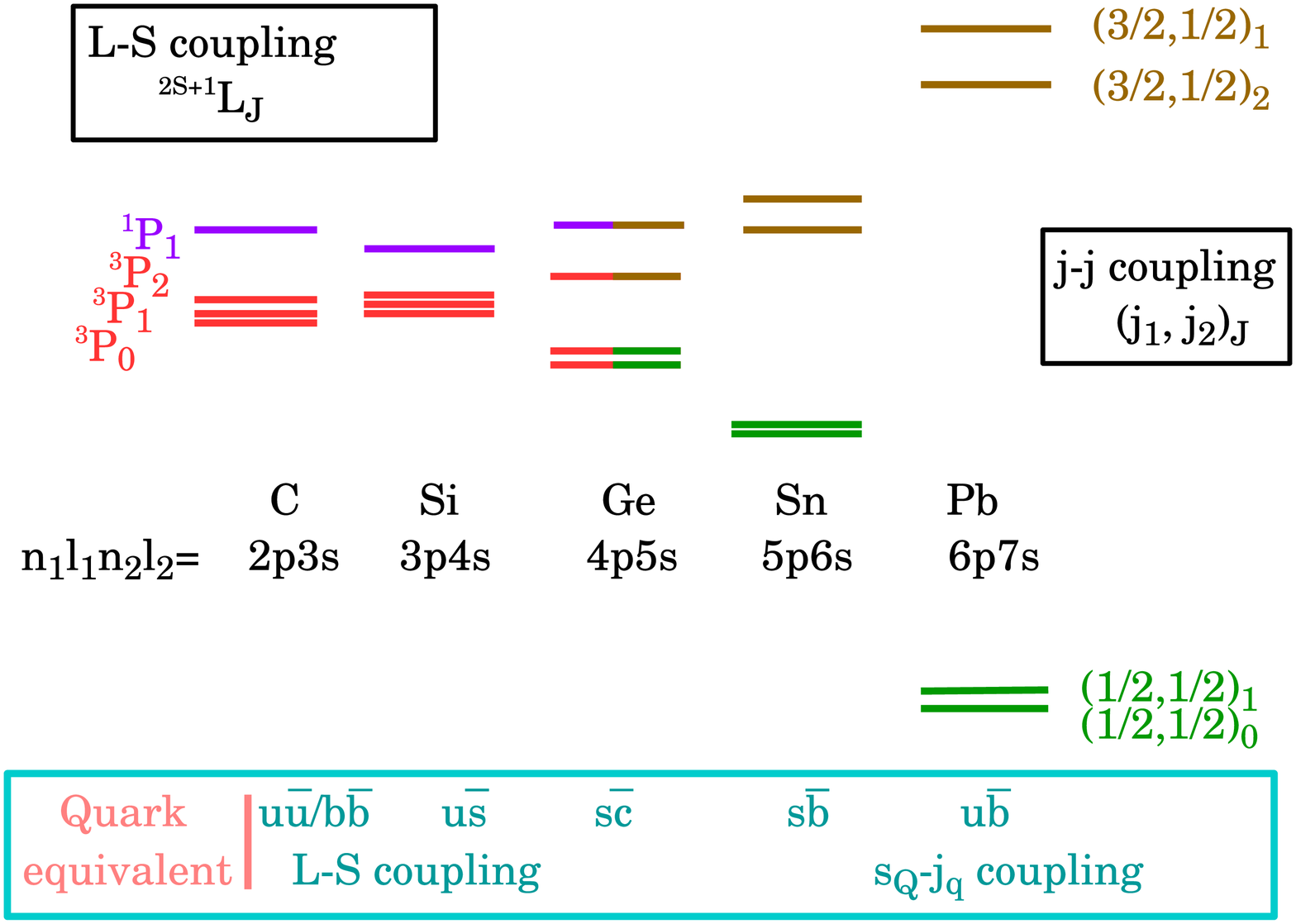}}
\end{center}
\end{minipage}\ \ \ 
\begin{minipage}{0.45\textwidth}
\begin{center}
\caption{\label{fig:CtoPb}
Elements in group 14 (old IV) of the periodic table have two electrons out of a closed shell as indicated. Carbon is a classic example in which these two electrons undergo Russell-Saunders ${\bf L}$--${\bf S}$ coupling just as in heavy quarkonium. Lead on the other hand shows very clear ${\bf j}$--${\bf j}$ coupling as in the heavy-light mesons. Intermediate elements nicely show the evolution between the two extreme cases.\\  Our meson calculations will likewise evolve from pure ${\bf L}$--${\bf S}$ to pure ${\bf j}\cdot {\bf j}$ coupling as function of the mass--difference between the quark and the antiquark (see figure~\ref{Fig-mixing-angle-s} below).
} 
\end{center}
\end{minipage}
\end{figure}

This is analogous to the ${\bf j}$--${\bf j}$ coupling in atomic physics and we dedicate figure~\ref{fig:CtoPb} to remind the reader of the transition, through group 14 (formerly, group IV) with two electrons outside closed spherical subshells, from Carbon to Lead, between quite pure Russell-Saunders ${\bf L}$--${\bf S}$ coupling for Carbon to quite pure ${\bf j}$--${\bf j}$ coupling for Lead.

It is instructive to illustrate these features by means of the shell model, in which the potential is given by $V=\sum V^{(i)}_{\rm central}+V^{ee}_{\rm residual}+V^{LS}$, with the residual electron-electron and spin-orbit interactions being written respectively as    
\begin{equation}
V^{ee}_{\rm residual}=\alpha\left( \sum_{i<j}^Z \frac{1}{\arrowvert {\bf r}_i-{\bf r}_j\arrowvert}
- \bigg< \sum_{i<j}^Z  \frac{1}{\arrowvert {\bf r}_i-{\bf r}_j\arrowvert} \bigg> \right)
\end{equation}
and
\begin{equation}
V^{LS} = \frac{1}{2m_e^2} \frac{1}{r} \frac{dV^{\rm central}}{dr} {\bf S}\cdot {\bf L}. 
\end{equation}
The ${\bf L}$--${\bf S}$ coupling is appropriate when the contribution coming from the residual electron-electron interaction (after subtracting the central part) dominates over the $S$-dependent spin-orbit interaction. Therefore, the remaining contribution does dot depend on $S$ so that it is a good quantum number.


Returning to the (infinitely--) heavy--light system~\cite{Rosner:1985dx}, since we know that $s_Q$ will be a good quantum number because the dominant term in the QCD Lagrangian is the spin--independent $m_Q\bar{\Psi}\Psi$, the mixing angle in the ${\bf L}$--${\bf S}$ basis can be exactly calculated.
In standard angular momentum notation,
\begin{equation}
\arrowvert \left( (L s_q)j_q s_Q \right) JM \rangle = \sum_S \arrowvert \left(L(s_q s_Q)S\right)JM\rangle \cdot
\langle (L(s_q s_Q)S)J\arrowvert ((L s_q)j_q s_Q)J\rangle
\end{equation}
that effects the change of basis in terms of a recoupling coefficient. This can be substituted by a Racah coefficient or a Wigner 6j coefficient, that for the problem at hand is
\begin{equation}
\bigg\langle \left (1\left(\frac{1}{2} \frac{1}{2}\right)S\right) 1 \bigg\arrowvert \left(\left(1 \frac{1}{2} \right)j_q \frac{1}{2} \right)1\bigg\rangle = 
(-1)^{1+1/2+1/2+1} \sqrt{2j_q+1} \sqrt{2S+1}
\left\{ 
\begin{array}{ccc}
$$1$$             & $$\frac{1}{2}$$ & $$j_q$$ \\
$$\frac{1}{2}$$   &   $$1$$         & $$S$$
\end{array}
\right\}\ .
\end{equation}
Evaluating the 6j coefficients finally leads to the rotation matrix
\begin{equation} \label{jqrotation}
\left( 
\begin{array}{c}
$$j_q=\frac{1}{2}$$ \\  $$j_q=\frac{3}{2}$$
\end{array}
\right)_{J=1} = 
\left( 
\begin{array}{cc}
$$\sqrt{\frac{2}{3}}$$ & $$-\sqrt{\frac{1}{3}}$$ \\
$$\sqrt{\frac{1}{3}}$$ & $$\sqrt{\frac{2}{3}}$$ 
\end{array}
\right)
\left( 
\begin{array}{c}
$$S=1$$ \\  $$S=0$$
\end{array}
\right)_{J=1}
\end{equation}
so that the two extreme basis for mesons are separated by a rotation angle 
\begin{equation} \label{maxangle}
\theta_P^{\rm max}=\arccos\left(\sqrt{\frac{2}{3}} \right)\simeq 35.3^{\rm o}\ . 
\end{equation}
Knowing this value exactly will come handy as a later check of the numerics.

Finally, light quarks deserve a specific comment. Though $m_d\gg m_u$, they are both much smaller than the QCD scale, $m_d, m_u \ll 1$ GeV. This causes isospin to be an approximate symmetry, and though for $u\bar{d}$ and $d\bar{u}$ mesons $C$ is not a good symmetry, it can be substituted for the approximate $G$-parity, that for a quark--antiquark system is $G:=C(-1)^I=(-1)^{L+S+I}$, with $I$ the isospin of the state.
In consequence, $S$ and $L$ are once more good quantum numbers (to fix the external $P$ and $G$) and these light mesons have a $q\bar{q}$ component that must be in the ${\bf L}$--${\bf S}$ basis.

In conclusion, when the quark and antiquark flavors are equal ($m_f=m_{f'}$), or when both are very small, the mixing angle vanishes. And when one of them is infinitely heavy but the other one is held fixed, the mixing angle takes the value $\arccos\left(\sqrt{\frac{2}{3}}\right)$. For intermediate cases, we will resort to an extraction from the computer code data.

In comparing to the literature, we need to take note that some authors use the opposite convention to the mixing angle, ordering the $LS$ basis by lowest $S$-spin instead of lowest mass; that is, instead of $(S=1,S=0)$ as in Eq.~(\ref{jqrotation}), they employ $(S=0,S=1)$. 
The two choices of angle are then complementary,   
$\theta_{P,\rm \ comp} = \frac{\pi}{2}-\theta_{P}$. Our choice makes the natural interval for the mixing angle be $[0,35.3^{\rm o}]$ and the complementary one $[54.7^{\rm o},90^{\rm o}]$. Other conventions still take $-\theta_{P,\rm \ comp}$. We find our choice the preferable one on the grounds of simple interpretation.
 
\section{Hamiltonian field theory formalism}
\label{sec:Formalism}

\subsection{Simplified Hamiltonian} \label{H}

In principle, one would like to solve the meson spectrum directly from the QCD Hamiltonian.
Its Coulomb gauge formulation~\cite{Christ:1980ku} has the advantage that one can construct the Fock space of possible hadrons directly from quarks, antiquarks and physical transverse gluon.
The disadvantage is a very difficult interaction kernel that depends on the fields (and, as in any equal--time Hamiltonian approach, a nontrivial boost operator that makes changes of reference frame all but intractable~\cite{Rocha:2009xq}).
For what is worth, we quote once again its exact form before proceeding to a sensible approximation:

\begin{equation}
	H_{QCD} = H_{q} + H_{g} + H_{qg} + H_{C}, 
\label{H_QCD}
\end{equation}
where 
\begin{eqnarray}
	H_{q} & = & \int d\mathbf{x} \Psi^{\dagger}\left(\mathbf{x}\right) \left[-i \mat{\alpha} \cdot \mat{\nabla} + \beta m\right] \Psi\left(\mathbf{x}\right), 
	\nonumber \\
	H_{g} & = & \frac{1}{2} \int d\mathbf{x} \left[\mathcal{J}^{-1} \mat{\Pi}^{a}\left(\textbf{x}\right) \cdot \mathcal{J} \mat{\Pi}^{a}(\textbf{x}) + \mathbf{B}^{a}(\textbf{x}) \cdot \mathbf{B}^{a}(\textbf{x})\right], 
	\nonumber \\
H_{qg} & = & g \int d\mathbf{x} \mathbf{J}^{a}\left(\textbf{x}\right) \cdot \mathbf{A}^{a}(\textbf{x}), 
\nonumber \\
	H_{C} & = & \frac{g^{2}}{2} \int d\mathbf{x} d\mathbf{y} \rho^{a}\left(\textbf{x}\right) \mathcal{J}^{-1} K^{ab}\left(\mathbf{x},\mathbf{y}\right) \mathcal{J} \rho^{b}\left(\textbf{y}\right).
	\label{H_QCD2}
\end{eqnarray}
There, $\Psi$ and $m$ are the current quark field and mass; $\mathbf{A}^{a}$ $(a=1,2,\ldots,8)$ are the Coulomb--gauge transverse gluon fields satisfying  $\mat{\nabla} \cdot \mathbf{A}^{a} = 0$; $g$ is the coupling constant; $\mat{\Pi}^{a}$ are the conjugate fields; $\mathbf{B}^{a}$ are the chromomagnetic fields
\begin{equation}
	\mathbf{B}^{a} = \mat{\nabla} \times \mathbf{A}^{a} + \frac{1}{2} g f^{abc} \mathbf{A}^{b} \times \mathbf{A}^{c};
	\label{Bfield}
\end{equation}
and the color densities $\rho^{a}$ and quark color currents $\mathbf{J}^{a}$ are given by 
\begin{eqnarray}
  	\rho^{a}(\textbf{x}) & = & \Psi^{\dagger}\left(\mathbf{x}\right) T^{a} \Psi\left(\mathbf{x}\right) + f^{abc} \mathbf{A}^{b}\left(\mathbf{x}\right) \cdot \mat{\Pi}^{c}\left(\mathbf{x}\right), \nonumber \\
	\mathbf{J}^{a} & = & \Psi^{\dagger}\left(\mathbf{x}\right) \mat{\alpha} T^{a} \Psi\left(\mathbf{x}\right), 
	\label{color_dens_curr}
\end{eqnarray}
with $T^{a} = \lambda / 2$ and $f^{abc}$ being the $SU_{c}(3)$ generators and structure constants, respectively. 

The factor $\mathcal{J}$ can be recognized as the Faddeev-Popov determinant and is defined as 
\begin{eqnarray}
	\mathcal{J} = \det\left(\mat{\nabla} \cdot \mathbf{D} \right),
	\label{FP_det}
\end{eqnarray}
where $\mathbf{D}$ is the covariant derivative in adjoint representation, $\mathbf{D}^{ab} = \delta^{ab} \mat{\nabla} - g f^{abc} \mathbf{A}^{c}$.  

Finally, the kernel $K^{ab}\left(\mathbf{x}, \mathbf{y}\right)$ in $H_{C}$ represents the instantaneous non-Abelian Coulomb interaction
\begin{eqnarray}
  	K^{ab}\left(\mathbf{x}, \mathbf{y}\right) = \left\langle \mathbf{x}, a \vert (\mat{\nabla} \cdot \mathbf{D})^{-1}( - \nabla^{2}) (\mat{\nabla} \cdot \mathbf{D})^{-1} \vert \mathbf{y}, b \right\rangle .
\label{Kernel1}
\end{eqnarray}
The nonlinear kernel and Faddeev-Popov determinant make the Hamiltonian of QCD in Coulomb gauge~\cite{Christ:1980ku} notoriously difficult to handle, one of the reasons why the Hamiltonian method is usually treated only in simplified terms~\cite{Feuchter:2004mk,Reinhardt:2004mm,Szczepaniak:2005xi}.

This work  addresses axial--vector mesons with $J^P=1^+$; but two transverse gluons, by Landau--Yang's theorem, cannot form a state of $J=1$; therefore, the term $H_{g}$ would start contributing only in three--particle configurations such as hybrid mesons~\cite{LlanesEstrada:2000hj} or three--gluon oddballs~\cite{LlanesEstrada:2005jf}. We do not need to discuss it in this paper, as the philosophy of the quasiparticle gap makes those configurations heavier than $q\bar{q}$ (which is supported by the calculations in those references).

To achieve a tractable model, we simplify  the remaining interaction terms $H_C$ and $H_{qg}$, replacing them by classical interactions. 
The Coulomb interaction is substituted by the following longitudinal Coulomb potential:
\begin{eqnarray}
  	H_{C} \longrightarrow V_{C} = -\frac{1}{2} \int d\mathbf{x} d\mathbf{y} \rho^{a}\left(\mathbf{x}\right) \hat{V}\left(\vert\mathbf{x} - \mathbf{y}\vert\right) \rho^{a}\left(\mathbf{y}\right),
  	\label{Coul_Pot1}
\end{eqnarray}
with a confining potential in momentum space derived from the Yang--Mills dynamics~\cite{Szczepaniak:2001rg}, 
\begin{eqnarray}
	V \left( p \right) = 
\begin{cases} 
	\left(-12.25 \frac{ m_g^{1.93}}{p^{3.93}} \right), & \mbox{for } p < m_g, \\ 
-\frac{8.07}{p^2} \frac{\ln{\left( \frac{p^2}{m_g ^2} + 0.82 \right)^{-0.62}}}{\ln{\left( \frac{p^2}{m_g ^2} + 1.41 \right)^{0.8}}}, & \mbox{for }  p > m_g.  \end{cases}
	  	\label{Coul_Pot2}
\end{eqnarray}
The parameter $m_g$ determines the scale of the model, and it is set to $m_g \approx 600 $ MeV.
This is in accordance with the obtention of such reasonable Cornell potential from gluodynamics~\cite{Szczepaniak:2001rg} and trades off the cutoff or equivalent regulator used to quantize it~\cite{Szczepaniak:2005xi}.

The coupling between quarks and transverse gluons $H_{qg}$ appears at second order in a diagrammatic expansion since the gluon has to be produced and absorbed. Having the structure $\vec{\alpha}\cdot\vec{\alpha}$ in spinor space, it is important to properly describe hyperfine splittings in the spectrum.
We again approximate this second order interaction by a classical transverse hyperfine potential $V_T$, 
\begin{eqnarray}
V_{T} & = \frac{1}{2} \int d\mathbf{x} \:d\mathbf{y} J_{i}^{a}\left(\textbf{x}\right) \hat{U}_{ij}\left(\mathbf{x}, \mathbf{y}\right) J^{a}_{j}(\textbf{y}), 
\label{V_T}
\end{eqnarray}
where the kernel $\hat{U}_{ij}$ inherits the transversality of the propagated physical gluons that have been eliminated,
\begin{eqnarray}
\hat{U}_{ij} \left(\mathbf{x}, \mathbf{y}\right) = \left(\delta_{ij} - \frac{\nabla_{i} \nabla_{j}}{\mat{\nabla}^{2}}\right)_{\mathbf{x}} \hat{U}\left(\vert \mathbf{x} - \mathbf{y} \vert\right).
\label{U_ij}
\end{eqnarray}
We choose $\hat{U}$ to be a Yukawa-type potential representing the exchange of a constituent gluon with dynamical mass $m_g$; in momentum space it is defined by 
\begin{eqnarray}
	U \left( p \right) = C_h\begin{cases} 
	(- 24.57) \frac{1}{p^2 + m_g ^2}, & \mbox{for } p < m_g, \\ 
- \frac{8.07}{p^2} \frac{\ln{\left( \frac{p^2}{m_g ^2} + 0.82 \right)^{-0.62}}}{\ln{\left( \frac{p^2}{m_g ^2} + 1.41 \right)^{0.8}}}, & \mbox{for }  p > m_g.  \end{cases}
	  	\label{Yuk_pot}
\end{eqnarray}
The constant $C_h$ is left as a free model parameter that controls the global strength of this potential with respect to the longitudinal one. The factor $-24.57$ is not a parameter, instead it is fixed by matching the high and low momentum ranges at the scale $m_g$. 

Thus, the model parameters are $m_g$ (overall scale), $C_h$ (purely phenomenological, in the gauge theory it should be fixed by $m_g$ or equivalently $\alpha_s$), and the current quark masses $m_f$. 
The model has the same degrees of freedom and global symmetries as QCD so its multiplet structure is the same; it supports spontaneous chiral symmetry breaking as described next in subsection~\ref{subsec:gap}, unlike the constituent quark model; having a long--range potential, it supports radial--like excitations (unlike the Nambu--Jona--Lasinio model, that has no excited states); its wave equations are much simpler to solve than the covariant Dyson--Schwinger equations in Landau gauge, where radial excitations are not well understood either; and unlike in lattice gauge theory, the formulation is continuous and the rotation and chiral properties are manifest. 

On the down side, there is no known way to control its uncertainties with a counting; and because the boost operators are complicated in equal--time quantization, its usefulness is limited to spectroscopy, hadron structure (form factors, structure functions, etc.) are not naturally treated in this framework, since they require wavefunctions in different reference frames.

\subsection{Quark gap equation}\label{subsec:gap}
The first order of business is to obtain gapped quasiparticles so a truncation of the Fock space at the $q\bar{q}$ level makes sense. Here we briefly summarize the gap equation obtained with the Bogoliubov-Valatin (BV) variational method, in the spirit of many earlier works~\cite{Adler:1984ri}. 
We introduce a variational trial function, $\phi (\arrowvert\mathbf{k}\arrowvert) \equiv \phi_k$, i.e. the Bogoliubov angle.  It specifies the quark vacuum and one--body dispersion relation by minimization of the vacuum expectation value of the Hamiltonian, $\delta \langle \Omega \vert H \vert \Omega \rangle = 0$, where $ \vert \Omega \rangle $ is the quasiparticle (BCS) vacuum. Then, proceeding with the standard minimization procedure with the convention for the quasiparticle basis in~\cite{LlanesEstrada:2004wr}, we obtain the quark gap equation, 
\ben
k s_k - m_f c_k & = & \int_{0} ^{\infty} \frac{q^2}{6 \pi ^2} \left[ s_k c_q \left(V_1 + 2 W_0 \right) - s_q c_k \left(V_0 + U_0 \right)\right]  , 
\label{gap_eq}
\een
where the shorthand functions $s_k$ and $c_k$ are defined in terms of the Bogoliubov
angle and can be related to the running quark mass $m_q (k)$ as
\ben
s_k \equiv \sin{\phi_k} = \frac{m_q(k)}{E(k)},  \nonumber \\
c_k \equiv \cos{\phi_k} = \frac{k}{E(k)},
\label{sc_def}
\een
with $E(k) = \sqrt{M_q ^2 (k) + k^2}$. 
The functions $V_0, V_1, W_0 $ and $U_0$ represent angular integrals of the form
\be \label{ang_int}
F_n (k,q) \equiv \int_{-1} ^{1} dx \; x^n \; F(|\mathbf{k} - \mathbf{q}|), 
\ee
 with $x = \hat{k}\cdot \hat{q}$. The $V_n$ and $U_n$ functions in Eq.~(\ref{gap_eq}) are thus angular integrals of the longitudinal and transverse potentials, respectively. The $W$-function is also connected to $U$, being defined for convenience by
\be 
W(|\mathbf{k} - \mathbf{q}|) \equiv U(|\mathbf{k} - \mathbf{q}|) \frac{x (k^2 + q^2) - k q (1 + x^2) }{|\mathbf{k} - \mathbf{q}|^2}. 
\label{W_func}
\ee
In the following sections, we will also make use of the auxiliary function $Z$: 
\be 
Z(|\mathbf{k} - \mathbf{q}|) \equiv U(|\mathbf{k} - \mathbf{q}|) \frac{ 1 - x^2 }{|\mathbf{k} - \mathbf{q}|^2}. 
\label{Z_func}
\ee

\begin{figure}
\begin{minipage}{0.45\textwidth}
\begin{center}
\includegraphics[width=\textwidth]{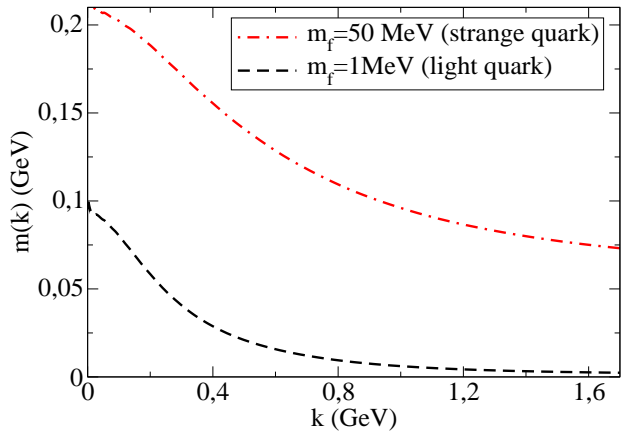}
\end{center}
\end{minipage}\ \ \ 
\begin{minipage}{0.45\textwidth}
\begin{center}
\caption{\label{fig:massgap}
Example gap functions $m(k)$ for $m_g=0.6$ GeV, $C_h=0.7$ and quark masses $m_u=1$ MeV, $m_s=50$ MeV at a high scale. The running masses increase from right (current) to left (constituent masses, respectively). }
\end{center}
\end{minipage}
\end{figure}

The gap equation~(\ref{gap_eq}) needs to be numerically solved, which we do by iteration with a Newton--like method (employing a linearization in the separation between the initial guess and the actual solution). A typical outcome is shown in figure~\ref{fig:massgap}.

\subsection{Meson spectrum: TDA equation of motion}

Once the 1--body problem has been variationally dealt with, we can interpret the mesonic states as excited bound states of quasiparticles. Safe for the pion, as a Goldstone boson, the Tamm-Dancoff (TDA) approximation~\cite{LlanesEstrada:2001kr} is appropriate.
First, let us deploy the more difficult TDA equation for an open-flavor meson in the state $| \Psi ^{nJP} \rangle $ with total angular momentum $J$, parity $P$ and radial quantum number $n$, 
\begin{eqnarray}
  	\langle \Psi^{nJP} \vert \left[H, Q^{\dagger}_{nJP} \right] \vert \Omega \rangle = \left(E_{nJP} - E_{0}\right) \langle \Psi^{nJP} \vert Q^{\dagger}_{nJP} \vert \Omega \rangle ; 
	\label{TDA_eq}
\end{eqnarray}
$Q^{\dagger}_{nJP}$ is the meson creation operator
\begin{eqnarray}
	Q^{\dagger}_{nJP} \equiv \sum_{\alpha \beta} \int\frac{d\mathbf{k}}{\left(2\pi\right)^{3}} \Psi^{nJP}_{\alpha \beta}\left(\mathbf{k}\right) B^{\dagger}_{\alpha}\left(\mathbf{k}\right) D^{\dagger}_{\beta}\left(-\mathbf{k}\right),
\label{meson_op}
\end{eqnarray}
with $B^{\dagger}_{\alpha}$ and $ D^{\dagger}_{\beta}$ the quasiparticle operators for the quark and antiquark, $\alpha, \beta $ denoting spin projections over $\bf k$ (we have omitted the color indices), and $\Psi^{nJP}_{\alpha \beta}$ the corresponding wave function.

Making use of conventional techniques, the commutators in left-hand side of  Eq.~(\ref{TDA_eq}) can be evaluated after normal ordering with respect to the BCS vacuum, and the projected equation for the wave function can be obtained. We employ the ${\bf L}$--${\bf S}$ basis
\begin{eqnarray}
	\Psi^{nJP}_{\alpha \beta}\left(\mathbf{k}\right) = \sum_{L S m_{L} m_{S}} \langle L, m_{L}, S, m_{S} \vert J, m_{J} \rangle \left(-1\right)^{\frac{1}{2} + \beta} \left\langle \frac{1}{2}, \alpha, \frac{1}{2}, -\beta \biggr\vert S, m_{S} \right\rangle Y_{L}^{m_{L}}\left( \hat{k} \right) \Psi^{nJP}_{LS}\left(k \right), 
\label{wave_func}
\end{eqnarray}
where 
$\Psi^{nJP}_{LS}\left(k\right)$ is the radial wave function. The equation for each of these components is then 
\begin{eqnarray}
 \left(M_{nJP} - \epsilon_{k}^{f} - \epsilon_{k}^{f'} \right) \Psi^{nJP}_{LS}\left(k\right) = \sum_{\Lambda \Sigma} \int\limits_{0}^{\infty} \frac{q^{2} dq}{12 \pi^{2}} \; K^{JP; f f'}_{L S ; \Lambda \Sigma}\left(k, q\right) \Psi^{nJP}_{\Lambda \Sigma}\left(q\right), 
\label{TDA_eq_part_wav}
\end{eqnarray}
where $M_{nJP} \equiv E_{nJP} - E_0$ is the mass of the meson state (if only one ${\bf L}$--${\bf S}$ component contributes) or a matrix (if more than one is coupled to the same $J^P$).
$\epsilon_{k } ^{f} $ is the self--energy of the quasiparticle with flavor $f$ (noticing that there is one gap angle for each quasiparticle), given by
\be 
\epsilon _{k } ^{f} = m _{f} s_{k} ^{f} + k c_{k} ^{f} - \int_{0} ^{\infty} \frac{q^2}{6 \pi ^2} \left[ s_{k }^{f} s_{q}^{f} \left(V_0 + 2 U_0 \right) + c_{k}^{f} c_{q}^{f} \left(V_1 + W_0 \right)\right]; 
\label{self_en}
\ee
and that needs to be regulated. It is formally infinite through the confining potential kernels $V_0$, $V_1$, but a Ward identity from global color symmetry guarantees the cancellation~\cite{Bicudo:1989si} of that infinity with the one coming from the two--body kernel (which checks all the relative factors in the computer code).

That kernel $K^{JP; f f'}_{L S ; \Lambda \Sigma}\left(k, q\right)$, coupling different orbital and spin states, is   given by
\begin{eqnarray}
	\nonumber K^{JP; f f'}_{L S ; \Lambda \Sigma}\left(k, q\right) &= & \frac{2}{\pi \left(2J + 1\right)} \sum_{m_{\Lambda} m_{\Sigma} m_{J} m_{L} m_{S}} \langle J, m_{J} \vert L, m_{L}, S, m_{S} \rangle \langle \Lambda, m_{\Lambda}, \Sigma, m_{\Sigma} \vert J, m_{J} \rangle \int d\Omega_{k} d\Omega_{q} Y_{L}^{*m_{L}}\left(\mathbf{k}\right)  Y_{\Lambda}^{m_{\Lambda}}\left(\mathbf{q}\right)  \\
	&  & \times \sum_{\gamma \delta \alpha \beta} \left(-1\right)^{1 + \beta + \gamma} G^{f f'}_{\alpha \beta \gamma \delta }\left(k, q\right) \left\langle S, m_{S} \biggr\vert \frac{1}{2}, \alpha, \frac{1}{2}, -\beta\right\rangle \left\langle \frac{1}{2}, \delta, \frac{1}{2}, -\gamma \biggr\vert \Sigma, m_{\Sigma} \right\rangle .
\label{Kernel}
\end{eqnarray}
The function $G^{\alpha \beta}_{ \gamma \delta }\left(k, q\right)$ in Eq.~(\ref{Kernel}) is defined as 
\begin{eqnarray}
G^{f f'}_{\alpha \beta \gamma \delta }\left(k, q\right) \equiv V\left(\vert\mathbf{k} - \mathbf{q}\vert\right) h^{f f'}_{\alpha \beta \gamma \delta }\left(k, q\right) - U\left(\vert\mathbf{k} - \mathbf{q}\vert\right) t^{f f'}_{\alpha \beta \gamma \delta }\left(k, q\right),
\label{G_func}
\end{eqnarray}
and carries dependence on the Bogoliubov angle coming from the quasiparticle basis through the functions $h^{f f'}_{\alpha \beta \gamma \delta }$ and $t^{f f'}_{\alpha \beta \gamma \delta }$, 
\begin{eqnarray}
	\nonumber h^{f f'}_{\alpha \beta \gamma \delta }\left(k, q\right) &= & \frac{1}{4} \left[a_{5} \:g_{ \gamma \beta}(\hat{\mathbf{k}}, \hat{\mathbf{q}}) \:\delta_{\alpha \delta} + a_{8} \:g_{ \gamma \beta}(\hat{\mathbf{k}}, \hat{\mathbf{q}}) \:g_{\alpha \delta}(\hat{\mathbf{q}}, \hat{\mathbf{k}}) + a_{7} \:\delta_{ \gamma \beta} \:\delta_{\alpha \delta} + a_{6} \:\delta_{ \gamma \beta} \:g_{\alpha \delta}(\hat{\mathbf{q}}, \hat{\mathbf{k}})\right], \\
	\nonumber t^{f f'}_{\alpha \beta \gamma \delta } \left(k, q\right) & = & - \frac{1}{4} \left[a_{1} b_{i \alpha \delta} ^L (\hat{\mathbf{k}}) b_{i \gamma \beta} ^R (\hat{\mathbf{k}})  + a_{3} b_{i \alpha \delta} ^L (\hat{\mathbf{k}}) b_{i \gamma \beta}^L (\hat{\mathbf{q}}) 
	+ a_{4} b_{i \alpha \delta} ^R (\hat{\mathbf{q}}) b_{ i \gamma \beta} ^R (\hat{\mathbf{k}}) + a_{2} b_{i \alpha \delta}^R (\hat{\mathbf{q}}) b_{i \gamma \beta} ^L (\hat{\mathbf{q}}) \right] \\
	\nonumber & & + \frac{1}{4 \left(\mathbf{k} - \mathbf{q}\right)^{2}}
	 \left[a_{1} \left(g_{\alpha \delta}\left(\hat{\mathbf{k}}, \hat{\mathbf{k}}\right) k - g_{\alpha \delta}\left(\hat{\mathbf{q}}, \hat{\mathbf{k}}\right) q\right) \left(g_{ \gamma \beta}\left(\hat{\mathbf{k}}, \hat{\mathbf{k}}\right) k - g_{ \gamma \beta}\left(\hat{\mathbf{k}}, \hat{\mathbf{q}}\right) q\right) +\right.\\
	\nonumber & & + a_{3} \left(g_{\alpha \delta}\left(\hat{\mathbf{k}}, \hat{\mathbf{k}}\right) k - g_{\alpha \delta}\left(\hat{\mathbf{q}}, \hat{\mathbf{k}}\right) q\right) \left(g_{ \gamma \beta}\left(\hat{\mathbf{k}}, \hat{\mathbf{q}}\right) k - g_{ \gamma \beta}\left(\hat{\mathbf{q}}, \hat{\mathbf{q}}\right) q\right) +\\
	\nonumber & & + a_{4} \left(g_{\alpha \delta}\left(\hat{\mathbf{q}}, \hat{\mathbf{k}}\right) k - g_{\alpha \delta}\left(\hat{\mathbf{q}}, \hat{\mathbf{q}}\right) q\right) \left(g_{ \gamma \beta}\left(\hat{\mathbf{k}}, \hat{\mathbf{k}}\right) k - g_{ \gamma \beta}\left(\hat{\mathbf{k}}, \hat{\mathbf{q}}\right) q\right) +\\
	& & \left. + a_{2} \left(g_{\alpha \delta}\left(\hat{\mathbf{q}}, \hat{\mathbf{k}}\right) k - g_{\alpha \delta}\left(\hat{\mathbf{q}}, \hat{\mathbf{q}}\right) q\right) \left(g_{ \gamma \beta}\left(\hat{\mathbf{k}}, \hat{\mathbf{q}}\right) k - g_{ \gamma \beta}\left(\hat{\mathbf{q}}, \hat{\mathbf{q}}\right) q\right)\right].
\label{h_t_func}
\end{eqnarray}
In these last expressions we have used the shorthands $g_{\alpha \beta}$ and $b_{i\alpha \beta} ^{L,R}$,
\ben
g_{\alpha \beta}\left(\hat{\mathbf{r}}, \hat{\mathbf{w}}\right)  & \equiv & \chi_{\alpha}^{\dagger} \mat{\sigma} \cdot \hat{\mathbf{r}} \mat{\sigma} \cdot \hat{\mathbf{w}}\chi_{\beta}, \nonumber \\
b_{i \alpha \beta} ^L \left(\hat{\mathbf{r}}\right) & \equiv &  \left(\chi_{\alpha}^{\dagger} \:\sigma_{i} \:\mat{\sigma} \cdot \hat{\mathbf{r}} \:\chi_{\beta}\right),
\nonumber \\ 
b_{i \alpha \beta} ^R \left(\hat{\mathbf{r}}\right) & \equiv & \left(\chi_{\alpha}^{\dagger}  \:\mat{\sigma} \cdot \hat{\mathbf{r}} \:\sigma_{i} \:\chi_{\beta}\right),
\label{g_b_func} 
\een
with $\chi_{\alpha} $ denoting Pauli spinors, and the coefficients $a_i$ that carry the gap angle dependence (and arise from the four--spinor products),
\begin{eqnarray}
	a_{1} &= & \sqrt{1 + s_{k}^f } \sqrt{1 + s_{k}^{f'} } \sqrt{1 - s_{q}^f } \sqrt{1 - s_{q}^{f'} },\label{eq:d17} \nonumber \\
	a_{2} &= & \sqrt{1 - s_{k}^f } \sqrt{1 - s_{k}^{f'} } \sqrt{1 + s_{q}^f } \sqrt{1 + s_{q}^{f'} }, \nonumber \\
	a_{3} &= &\sqrt{1 + s_{k}^f } \sqrt{1 - s_{k}^{f'} } \sqrt{1 - s_{q}^f } \sqrt{1 + s_{q}^{f'} }, \nonumber \\
	a_{4} &= & \sqrt{1 - s_{k}^f} \sqrt{1 + s_{k}^{f'} } \sqrt{1 + s_{q}^f} \sqrt{1 - s_{q}^{f'} }, \nonumber \\
	a_{5} &= & \sqrt{1 + s_{k}^f } \sqrt{1 - s_{k}^{f'} } \sqrt{1 + s_{q}^f } \sqrt{1 - s_{q}^{f'} }, \nonumber \\
	a_{6} &= & \sqrt{1 - s_{k}^f } \sqrt{1 + s_{k}^{f'} } \sqrt{1 - s_{q}^f } \sqrt{1 + s_{q}^{f'} }, \nonumber \\
	a_{7} &= & \sqrt{1 + s_{k}^f } \sqrt{1 +  s_{k}^{f'} } \sqrt{1 + s_{q}^f } \sqrt{1 + s_{q}^{f'} }, \nonumber \\
	a_{8} &= & \sqrt{1 - s_{k}^f } \sqrt{1 - s_{k}^{f'} } \sqrt{1 - s_{q}^f } \sqrt{1 - s_{q}^{f'} }.
	\label{a_coeff}
\end{eqnarray}
In turn, $s_{k(q)}^{f(f')} $ is the sine of the corresponding gap angle as given in Eq.~(\ref{sc_def}), obtained by solving the gap equation for the $f(f')$-th quasiparticle. It carries the dependence on the current quark mass and (for light quarks) on chiral symmetry breaking.

Application of the TDA equation to the meson spectrum with quantum states designated by $I^G (J^{PC})$, requires first an analytic computation of the corresponding kernel $K^{JP;f f'}_{L S ; \Lambda \Sigma}\left(k, q\right)$. 
A few of the lowest angular momentum kernels, assuming isospin symmetry 
(and omitting the $f$, $f'$ indices) are: 
  \begin{itemize}
	\item pseudoscalar ($0^{-+}$),
\begin{eqnarray}
        K^{0^-}_{0 0 ; 0 0}\left(k, q\right)  & = &   V_{1} \left(a_{5} + a_{6}\right) + V_{0} \left(a_{7} + a_{8}\right) + 2 U_{0} \left(a_{1} + a_{2}\right) 
			- 2 W_{0} \left(a_{3} + a_{4}\right);
	\label{K_PS}
\end{eqnarray}
(Actually, we employ an extended version of this equation using the Random Phase Approximation as described in~\cite{LlanesEstrada:1999uh} that respects chiral symmetry, guaranteeing that Goldstone's theorem is implemented and thus $m_\pi=0$ in the $m_q=0$ limit, but we eschew a detailed description because it defocuses our discussion of the axial vector mesons for which the TDA is sufficient.)
	
	\item vector ($1^{--}$),
\begin{eqnarray}
			K^{1^-}_{0 1 ; 0 1}\left(k, q\right)  & = &   \frac{1}{3} \left[3 V_{1} \left(a_{5} + a_{6}\right) + a_{8} \left(4 V_{2} - V_{0}\right) + 3 a_{7} V_{0} - 2 \left(a_{1} + a_{2}\right) U_{0} +\right. \nonumber \\
			& & \left.+ 2 \left(a_{3} + a_{4}\right) U_{1} + 2 q k \left(a_{3} + a_{4}\right) Z_{0} + 4 \left(a_{1} k^{2} + a_{2} q^{2}\right) Z_{0} \right];
	\label{K_V}
\end{eqnarray}
		
	\item axial ($1^{+-}$),
\begin{eqnarray}
			K^{1^+}_{1 0 ; 1 0}\left(k, q\right)  & = &   \left(a_{5} + a_{6}\right) V_{2} + \left(a_{7} + a_{8}\right) V_{1} + 2 \left(a_{1} + a_{2}\right) U_{1}
			 - 2 \left(a_{3} + a_{4}\right) W_{1};
	\label{K_PV1}
\end{eqnarray}
		
	\item axial ($1^{++}$),
\begin{eqnarray}
			K^{1^+}_{1 1 ; 1 1}\left(k, q\right)  & = &   \frac{1}{2} \left(V_{0} + V_{2}\right) \left(a_{5} + a_{6}\right) + \frac{1}{2} \left(U_{0} + U_{2} - 2 W_{1}\right) \left(a_{3} + a_{4}\right) \nonumber  \\
			& & + V_{1} \left(a_{7} + a_{8}\right) + Z_{1} \left(a_{1} k^{2} + a_{2} q^{2}\right) + Z_{0} \frac{1}{2} \left(k^{2} - q^{2}\right) \left(a_{4} - a_{3}\right).
	\label{K_PV2}
\end{eqnarray}

\end{itemize}
The parts proportional to the longitudinal Coulomb potential (all terms containing $V_i$) can be checked against prior literature, as are the entire pseudoscalar and vector kernels.
The longitudinal axial--vector kernel pieces  in Eqs.~(\ref{K_PV1}) and~(\ref{K_PV2}) coincide with those computed by~\cite{Ligterink:2003hd} (that corrected an error in the earlier evaluation of~\cite{LlanesEstrada:2001kr}, where the scalar and tensor kernels for the longitudinal potential can be found if needed).

\subsection{Nondiagonal TDA equation (for open flavor)}
\label{Off-diagonal}

Central to this work is the mixing of axial states with open flavor $f\neq f'$, 
$ \vert u\bar{s} \rangle$,  $\vert c\bar{u}\rangle$,  $\vert c\bar{s}\rangle$, etc.
in which case the quark and antiquark have different gap angles.
(In the case of hidden flavor $f=f'$, the gap angle is the same for both, and therefore the spectrum can be obtained by solving the eigenvalue problem in TDA equation with the kernels given by Eqs.~(\ref{K_PV1}) and (\ref{K_PV2}) in the ${\bf L}$--${\bf S}$ basis; this degenerate case is relegated, for the sake of expediency, to appendix~\ref{sec:equalflavor}.)

For states with open flavor we should expect the Coulomb gauge model to incorporate mixing, 
yielding non-vanishing off-diagonal elements of the Hamiltonian.  

The TDA equation given in Eq.~(\ref{TDA_eq_part_wav}) generalizes then to a coupled--channel problem given by 
\begin{eqnarray} \label{Mixeddiagonalization}
	 \left(M_{n1^+} - \epsilon_{k}^{f} - \epsilon_{k}^{f'}\right) 
	\left(\begin{array}{c}
 \Psi^{n1^+}_{10}\left(k\right) \\
 \Psi^{n1^+}_{11}\left(k\right) \\
\end{array}\right) = \int\limits_{0}^{\infty}  \frac{q^{2} dq}{12 \pi^{2}} \; 
\left(\begin{array}{cc}
K^{1^+}_{1 0 ; 1 0}\left(k, q\right) & K^{1^+}_{1 0 ; 1 1}\left(k, q\right) \\
K^{1^+}_{1 1 ; 1 0}\left(k, q\right) & K^{1^+}_{1 1 ; 1 1}\left(k, q\right) \\
\end{array}\right) 
\left(\begin{array}{c}
 \Psi^{n1^+}_{10}\left(q\right) \\
 \Psi^{n1^+}_{11}\left(q\right) \\
\end{array}\right), 
\label{TDA_eq_part_wav_Axial}
\end{eqnarray}
where the off-diagonal element $K^{1^+}_{1 0 ; 1 1}\left(k, q\right)$ of the kernel matrix is given by 
\begin{eqnarray}
			K^{1^+}_{1 0 ; 1 1}\left(k, q\right)  &  = &   \frac{1}{\sqrt{2}} \left\{ \left(V_{2} - V_{0}\right) \left(a_{5} - a_{6}\right) + \left[ U_{0} - U_{2} + Z_{0} \left( k^2 - q^2 \right) \right] \left(a_{3} - a_{4}\right) \right\}.
	\label{K_PV_OFF_DIAG}
\end{eqnarray}
We can exploit the symmetry of the TDA kernels under transposition and $\mathbf{k} \leftrightarrow \mathbf{q}$ exchange, to yield  $ K^{1^+}_{1 1 ; 1 0}\left(k,q\right) =  K^{1^+}_{1 0 ; 1 1}\left(q, k\right) $. 

An interesting check is to take both quasiparticles to have equal flavor.
In that case, $a_{4} = a_{3}$ and  $a_{5} = a_{6}$ as can be read off Eq.~(\ref{a_coeff}) setting $f=f'$ there. In that case the kernel $K$ in Eq.~(\ref{K_PV_OFF_DIAG}) vanishes
and the ${\bf L}$--${\bf S}$ basis diagonalizes the Hamiltonian. 
This is as advertised since only the diagonal elements $\langle 1^{++} \vert H \vert 1^{++} \rangle $ and $\langle 1^{+-} \vert H \vert 1^{+-} \rangle $ should be finite. The off-diagonal elements  $\langle 1^{++} \vert H \vert 1^{+-} \rangle $ and $\langle 1^{+-} \vert H \vert 1^{++} \rangle $ must then vanish due to $C$-parity becoming a good quantum number. 
(Taking this limit analytically and numerically, we obtain the same results as those described in Appendix~\ref{sec:equalflavor} for the equal--flavor case.)

In the general case when $a_3\neq a_4$, $a_5\neq a_6$, the solution of the integral eigenvalue problem  in Eq.~(\ref{TDA_eq_part_wav_Axial}) provides the masses of the mixed pseudovector states.

However, the formulation in Eq.~(\ref{K_PV_OFF_DIAG}) makes clear that the oft discussed mixing angle $\theta_P$ is, strictly speaking, insufficient to completely describe the $^3P_1$--$^1P_1$ mixing.

\begin{figure}
\begin{minipage}{0.45\textwidth}
\begin{center}
\includegraphics[width=\columnwidth]{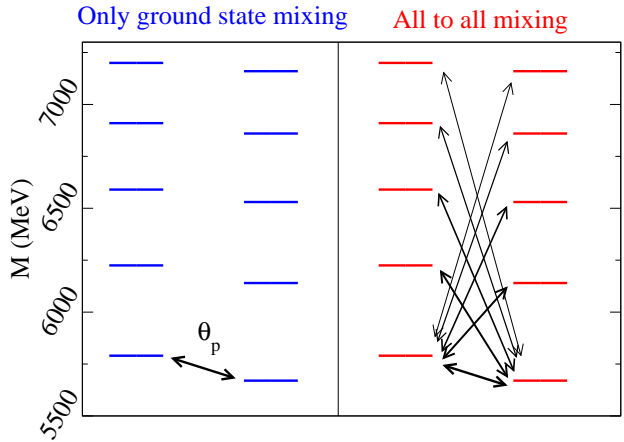}
\end{center}
\end{minipage}\ \ \ 
\begin{minipage}{0.45\textwidth}
\begin{center}
\caption{\label{fig:mixingtype} Whereas in the literature one often discusses the mixing angle between the ground state $^3P_1$ and $^1P_1$ state (left), a more general treatment as in Eq.~(\ref{Mixeddiagonalization}) allows for each of the radial excitations on the $^1P_1$ tower to mix with any of those in the $^3P_1$ suite, since the radial $q\bar{q}$ wavefunctions are orthonormal only within each of the two sets, but not across them. This more general mixing is depicted in the right panel. (The levels actually correspond to the axial $B_1$ mesons computed with the Coulomb gauge approach.)}
\end{center}
\end{minipage}
\end{figure}
As figure~\ref{fig:mixingtype} shows, the usual treatment in terms of only one mixing angle $\theta_P$ misses the fact that any of the $^3P_1$ states can mix with any of the $^1P_1$ levels. Naturally, the ground state mixes more strongly with the ground state. But our treatment actually allows for a full simultaneous diagonalization of the two towers of states to yield a unique $1^+$ spectrum for each flavor combination.

Nevertheless, we will loosely speak of the mixing angle $\theta_P$ extracting it phenomenologically from our resulting calculated spectrum. It should be clear though that the small mixing with excited states of the opposite ${\bf L}$--${\bf S}$ coupling causes probability leak to a wider Hilbert space. 
In a strict $2\times 2$ treatment as often done in phenomenological work, one needs to allow for $\theta_P$ to have a small imaginary part representing the leak in the reduced, ground state, space.

\section{Selected numerical spectrum for open--flavor $1^+$ mesons.}
\label{Results}

In this section we report on the calculated spectra for the axial-vector mesons with open flavor. 
They are obtained by solving the gap equation followed by the eigenvalue problem in Eq.~(\ref{TDA_eq_part_wav_Axial}),  including both the improved Cornell potential and the transverse hyperfine interaction whose kernel is a Yukawa-type potential, corresponding to the exchange of a constituent gluon with a dynamical mass $m_g $, as discussed in subsection~\ref{H}. The parameters used to obtain them are $m_g = 600 $ MeV,  $C_h = 0.7$ and the quark masses quoted in each table. All integrations have been cutoff at a scale $\Lambda = 6.0 $ GeV. 
There is little sensitivity to this cutoff at or above this scale, since the typical support of the wavefunction for the ground state mesons in each channel is a few hundred MeV, so the precise value of $\Lambda$ is of little consequence. The gluon mass $m_g$ was fixed from~\cite{Szczepaniak:2001rg,LlanesEstrada:2004wr} to obtain a reasonable Cornell potential, so it is ultimately tied to the charmonium spectrum through other works. $C_h$ is not a parameter directly relatable to QCD
and we use it to have sensible pseudoscalar--vector meson splittings across the quark--mass range (in combination with chiral symmetry breaking, that enhances the $\pi$--$\rho$ splitting). 
Because these few numbers are needed to obtain agreement with the basic pseudoscalar and vector mesons, the axial--vector computations are parameter--free. Nevertheless, we will show the dependence on the current quark mass which we believe is the most interesting dependence and our focus.

The current quark masses $m_f$ approximately corresponding to a physical flavor in the model approach and the constituent quark masses $\mathcal{M}_f = M_f (0) $ extracted from the gap equation are displayed in Table~\ref{TABLE-QUARKS}.

\begin{center}
\begin{table}[h!]
\caption{The current and constituent quark masses ($m_f$ and $\mathcal{M}_f = M_f (0)$, respectively). All quantities are given in MeV. }
\vskip1.5mm
\label{TABLE-QUARKS}
\begin{tabular}{c | c c }
\hline
\hline
Flavor & This approach   & Other related   \\
       & (Coulomb gauge QCD model) & estimates~\cite{LlanesEstrada:2004wr,Tanabashi:2018oca} \\
\hline
$m_u = m_d$ & 1  & 1.5-5.5
\\
$m_s $ & 50  & 70-120
\\
$m_c $ & 830  & 1000-1400
\\
$m_b $ & 3900  & 4000-4500
\\
\hline
$\mathcal{M}_u = \mathcal{M}_d$ & 97 & 200-340
\\
$\mathcal{M}_s $ & 208  & 450-500
\\
$\mathcal{M}_c $ & 1218  & 1500-1600
\\
$\mathcal{M}_b $ & 4436  & 4600-5100
\\
\hline
\hline
\end{tabular}
\end{table}
\end{center}

The input quark masses seem somewhat smaller than other estimates from quark models, but this field theory approach has a contribution from the quark self--energy that works to increase the meson masses in practice, so the constituent quark masses (and thus, their current masses too) need to be smaller to reasonably reproduce the basic pseudoscalar and vector mesons.

The ``constituent" quark mass is a model--dependent concept. In the quark model of Godfrey and Isgur~\cite{Godfrey:1985xj} or later similar approaches, the one--body part of the Hamiltonian is $\epsilon(k) = M + T(k)$ where $T(k)$ is an explicit function of $k$ that does not receive a contribution from the interaction potential. In field--theory approaches such as this Coulomb gauge model, the one--body part is given by Eq.~(\ref{self_en}), that, schematically and near zero momentum, takes the form $\epsilon(k) = M + T(k)+\int dq F[V(k,q)]$. The last contribution is positive for a potential attractive in the $q\overline{q}$ channel (note the sign in Eq.~(\ref{Coul_Pot2})) and accounts for the difference between the relatively light constituent masses in table~\ref{TABLE-QUARKS} and traditional nonrelativistic approaches. With a strong infrared confining interaction however, this integral is divergent by itself; only when used in the two--body equation~(\ref{TDA_eq_part_wav}), a cancellation with the negative two--body potential yields the finite meson masses. In summary, it is not surprising that the constituent quark masses are smaller than in more static constituent approaches. The constituent mass is still a useful concept marking the onset of spontaneous chiral symmetry breaking. Other relativistic approaches, however~\cite{Santopinto:2014opa}, obtain similarly light quark masses.

The BCS gap equation for the one--body problem and the TDA (or RPA, for the pseudoscalar) are solved in the same grid, which is important to aid with the numerical cancellation of the infinity in the self--energy and that in the two--body kernel. The resulting radial wavefunctions $\Psi^{nJP}_{LS}(\arrowvert {\bf k}\arrowvert)$ are expressed in that grid as $\Psi^{nJP}_{LS}(k_i)$, though we do not address them in this work.
(An alternative method employing a variational basis of a few bell--shaped functions instead of deltas at the $k_i$ points was put forward in \cite{Amor-Quiroz:2017jhs}.)

The physical $J=1$ states are linear combinations of the ${n}^3 P _1$ and ${n}^1 P _1$  basis states, and we can obtain them by considering the off-diagonal matrix element discussed in Sec.~\ref{Off-diagonal} relating these ${n}^3 P _1$ and ${n}^1 P _1$ states to the physical states $n P_1 $ and $n P_1 ^{\prime}$ [with the index prime $({}^{\prime})$ indicating the lowest and highest eigenvalues].

As already discussed, the mixing parameter is approximately a mixing angle. 
To compare with the literature, we can obtain such angle from a mock--theory in which, 
instead of Eq.~(\ref{Mixeddiagonalization}), the eigenvectors $n P_1 $ and $n P_1 ^{\prime}$ would stem from
diagonalization of a $2 \times 2$ mass matrix: this yields a relation between the mixing angle $\theta _P$ and the mass differences~\cite{Blundell:1995au}, 
\ben
\cos{2 \theta _{nP}} = \frac{M(n^1P_1) - M(n^3P_1)}{M(n P_1 ) - M(n P_1 ^{\prime})};
\label{mixingangle}
\een
as well the corresponding masses
\ben
M(n P_1 ) & = & M(n^1P_1) \cos^2{\theta _{nP}} + M(n^3P_1) \sin^2{\theta _{nP}} - [M(n^3P_1) - M(n^1P_1)] \frac{\sin^2{2 \theta _{nP}}}{2 \cos{2 \theta _{nP}}},  \nonumber \\
M(n P_1 ^{\prime}) & = & M(n^1P_1) \sin^2{\theta _{nP}} + M(n^3P_1) \cos^2{\theta _{nP}} + [M(n^3P_1) - M(n^1P_1)] \frac{\sin^2{2 \theta _{nP}}}{2 \cos{2 \theta _{nP}}} . 
\label{mixing}
\een
Table~\ref{TABLE-MESONS} summarizes the TDA masses of the lowest-lying $n P_1 $ and $n P_1 ^{\prime}$ mesons with open flavor; the energies of the $^3P_1$ and $^1P_1$ configurations in the absence of mixing; and the  
$1^{+}$--$1^{+'}$ 
mixing angles resulting from their comparison.

\begin{center}  
\begin{table}[h!]
\caption{TDA masses of lowest-lying unmixed
$1^1P_3$, $1^1P_1$,   as well as mixed (physical) 
$1 P_1 $ and $1 P_1 ^{\prime}$ mesons with open flavor,  and the 
$1^{+}$--$1^{+'}$ 
mixing angles. The TDA eigenvalue problem as well as the gap equation have been solved with the presence of an improved Cornell potential and a transverse hyperfine interaction, as discussed in subsection~\ref{H}. The masses are given in GeV and rounded off to the nearest 5 MeV after estimating the mixing angle. 
The experimental values of the masses $1 P_1 $ and $1 P_1 ^{\prime}$  states are given in the third column, when available~\cite{Tanabashi:2018oca}. 
The last column reports the mixing angle in the relativized model of Godfrey and Isgur taken from Godfrey and Isgur (GI)~\cite{Godfrey:1985xj} or Ferretti and Santopinto (FS)~\cite{Ferretti:2015rsa}. Whereas our mixing angle interpolates between $LS$ ($\theta_P=0$) and $jj$ ($\theta_P \simeq 35.3^{\circ}$) coupling values, for equal--flavor and very different flavored quarkonia respectively, the logic of $\theta_{\rm QM}$ in the literature is less clear, though some of their smaller values for the angle can perhaps be understood by the constituent mass being larger (so that the $jj$--type mixing is further away).
}
\vskip1.5mm
\label{TABLE-MESONS}
\begin{tabular}{c | c | c| c c c c |c|c}
\hline
Quark content $[q_f \bar{q_{f'}}]$   & 
  States   $(1P_1, 1P_1^{\prime})[I(J^P)]$ &  
Exp. mass (PDG)  &  
$1^{++}$ & $1^{+-}$ & 
$1P_1$ & 
$1P_1^{\prime}$
& $\theta_P$ & 
 $\theta^{\rm QM}_{\rm comp}=\frac{\pi}{2}-\theta_{\rm QM}$    
\\
\hline
$s\bar{u} / s\bar{d} $ & $ K_1(1270),  K_1(1400) [\frac{1}{2}(1^+)]$ &  1.272, 1.403 &  1.180 & 1.375  &  1.135  &    1.410  &  22.3$^{\circ}$ &  
-34$^{\circ}$ (GI)
\\
$c\bar{u} / c\bar{d} $ & $ D_1(2420), D_1(2430) [\frac{1}{2}(1^+)]$ & 
2.422, 2.423
 & 2.350  &  2.490  &   2.225  &    2.600   & 34.0$^{\circ}$ & 
25.7$^{\circ}$ (FS); 41$^{\circ}$ (GI) 
\\
$c\bar{s} $ & $ D_{s1}(2460), D_{s1}(2536) [0(1^+)]$ & 2.460, 2.536  & 2.420  & 2.515  &  2.350  & 2.580 & 33.0$^{\circ}$ & 
37.5$^{\circ}$ (FS); 44$^{\circ}$ (GI) 
\\
$b\bar{u} / b\bar{d} $ & $ B_1(5721),? [\frac{1}{2}(1^+)]$ & 5.726, ? & 5.665 &  5.790  &  5.535 &     5.905 &  35.0$^{\circ}$  & 
30.3$^{\circ}$ (FS); 43$^{\circ}$ (GI)  
\\
$b\bar{s} $ & $ B_{s1}(5830),? [0(1^+)]$ & 5.829, ?  &  5.725  &  5.810  &  5.645  &  5.890  &  34.8$^{\circ}$ & 
39.1$^{\circ}$ (FS); 45$^{\circ}$ (GI) 
\\
$b\bar{c} $ & $ ?, ? [0(1^+)]$ & ?, ?  & 6.595  &   6.610  &  6.580 &     6.620  & 33.4$^{\circ}$ &  
53$^{\circ}$ (GI) 
\\
\hline
\end{tabular}
\end{table}
\end{center}
The information in this Table~\ref{TABLE-MESONS} shows that the parameter--free prediction of the Coulomb--gauge kernels gets the spectrum approximately right but is not particularly accurate, with typical errors 100-200 MeV. We do not consider that it is worth fine--tuning it, since it is not an arbitrarily improvable approximation with a control parameter, and rather proceed to make some more general statements. 
Should one wish to identify the reasons of the discrepancies with the experimental spectrum, even realizing that global models of hadrons always carry an uncertainty barring parameter fine tuning for each subsystem of the meson spectrum, one should first start thinking that the potential
itself, in a truncated Hilbert space, can acquire chiral symmetry breaking pieces that change the spin splittings. Our four--vector interaction, borrowed directly from the QCD Lagrangian, needs to incorporate additional terms that require calculations in a self--consistent way (see~\cite{Alkofer:2008tt} for a lucid discussion in Landau gauge). Such pieces enhance the difference of spin couplings between light and heavy quarks and (presumably) improve the global fit. Further improvement likely require multiquark--explicit gluon states.

The one thing that can be examined, in an approach that simultaneously incorporates light--quark and heavy--quark symmetries as appropriate, is the dependence of the spectrum with the quark masses.
This is particularly interesting for the mixing angle, that depends quite strongly on $m_f-m_{f'}$.
To this purpose we dedicate Tables~\ref{TABLE-Axial-S},  and \ref{TABLE-Axial-D}.

They display the masses of the axial vector states, the $1^{++} - 1^{+-}$ mixing angles as a function of the current quark mass $m_{f'}$, at fixed value of  $m_f \equiv m_s, m_c$. 
(As a check, these calculations were carried out with a substantially larger number of points in the integral equation discretization than those in Table~\ref{TABLE-MESONS}. The difference in the eigenstates are not visible within our quoted 5 MeV precision.)

\begin{center}
\begin{table}[h!]
\caption{\emph{Axial vector mesons with at least one $s$--quark}. \\ Masses of the axial vector states $^3P_1$, $^1P_1$,  $1 P_1 $ and $1 P_1 ^{\prime}$ (and of the first radially excited doublet  $2 P_1 $ and $2 P_1 ^{\prime}$)  and the 
$^3P_1 - ^1P_1$ mixing angle for the ground state, as a function of the current quark mass $m_{f'}$, at fixed value of  $m_f  = 50 $ MeV ($\equiv m_s$). All meson masses in GeV. The (orientative) physical points are highlighted in boldface. The experimental values of the masses from the PDG are given when available.} 
\vskip1.5mm
\label{TABLE-Axial-S}
\begin{tabular}{c | c c c c c| c c  }
\hline
$m_{f'}$ & $^3P_1$ & $^1P_1$ &  $1 P_1 $  &   $1 P_1 ^{\prime}$ & $\theta _P$ &  $2 P_1 $ &  $2 P_1 ^{\prime}$ \\
\hline
0.000 & 1.175 & 1.375 & 1.130 & 1.410 & 22.7$^{\circ}$ & 1.830 & 1.995 \\
\hline
{\bf 0.001} & {\bf 1.180} & {\bf 1.375} & {\bf 1.135} & {\bf 1.410} & {\bf 22.2$^{\circ}$}
 & {\bf 1.840} & {\bf 2.000}  \\ 
 {\bf State} & & & $K_1(1270)$ & $K_1(1400)$ & & $K_1(1650)$ &  \\ 
 $\mathbf{m_{exp}}$  & & & 1.272(7) & 1.403(7) & & 1.672(50) &  \\ 
\hline
0.005 & 1.190 & 1.380 & 1.160 & 1.405 & 20.2$^{\circ}$ & 1.855 & 2.000\\
0.010 & 1.205 & 1.385 & 1.180 & 1.405 & 17.6$^{\circ}$ & 1.875 & 2.000\\
0.015 & 1.215 & 1.390 & 1.200 & 1.405 & 15.7$^{\circ}$ & 1.890 & 2.005\\
0.020 & 1.230 & 1.400 & 1.215 & 1.405 & 13.3$^{\circ}$ & 1.905 & 2.010\\
0.025 & 1.240 & 1.405 & 1.235 & 1.410 & 10.6$^{\circ}$ & 1.920 & 2.015\\
0.030 & 1.250 & 1.415 & 1.245 & 1.415 & 7.7$^{\circ}$ & 1.930 & 2.020\\
0.035 & 1.260 & 1.420 & 1.260 & 1.420 & 6.4$^{\circ}$ & 1.940 & 2.030\\
0.040 & 1.270 & 1.430 & 1.270 & 1.430 & 4.6$^{\circ}$ & 1.950 & 2.035\\
0.045 & 1.285 & 1.435 & 1.285 & 1.435 & 0$^{\circ}$  & 1.960 & 2.045 \\
\hline
{\bf 0.050} & {\bf 1.295} & {\bf 1.445} & {\bf 1.295} & {\bf 1.445} & {\bf 0$^{\circ}$}  & {\bf 1.970} & {\bf 2.050}  \\
 {\bf State} & $f_1(1420)$ & $h_1(1415)$ & & & &  &  \\ 
 $\mathbf{m_{exp}}$  & $1.426(1) $&$ 1.416 (8)$& &  & & &  \\ 
\hline
0.055 & 1.305 & 1.450 & 1.305 & 1.450 & 0$^{\circ}$  & 1.980 & 2.060 \\
0.060 & 1.315 & 1.460 & 1.315 & 1.460 & 4.7$^{\circ}$ & 1.985 & 2.065 \\
0.065 & 1.325 & 1.460 & 1.320 & 1.470 & 6.7$^{\circ}$ & 1.995 & 2.075\\
0.070 & 1.335 & 1.475 & 1.330 & 1.480 & 7.5$^{\circ}$ & 2.000 & 2.080\\
0.075 & 1.345 & 1.485 & 1.340 & 1.485 & 7.5$^{\circ}$ & 2.010 & 2.090\\
0.080 & 1.355 & 1.495 & 1.350 & 1.495 & 9.5$^{\circ}$ & 2.020 & 2.095\\
0.085 & 1.365 & 1.500 & 1.355 & 1.505 & 11.1$^{\circ}$ & 2.025 & 2.105\\
0.090 & 1.370 & 1.510 & 1.365 & 1.515 & 11.6$^{\circ}$ & 2.030 & 2.115\\
0.095 & 1.380 & 1.515 & 1.375 & 1.525 & 12.5$^{\circ}$ & 2.040 & 2.120\\
0.100 & 1.390 & 1.525 & 1.380 & 1.535 & 14.1$^{\circ}$ & 2.045 & 2.130\\
\hline
\end{tabular}
\end{table}
\end{center}

Table~\ref{TABLE-Axial-D} provides the masses and mixing angle of the first axial doublet with one charm quark, as function of the mass of the antiquark $m_{f'}$.

\begin{center}
\begin{table}[h!]
\caption{
\emph{Axial vector mesons with at least one $c$--quark}.\\ 
Masses of the axial vector states $^3P_1$, $^1P_1$, $1_L ^{+}$,  $1_H ^{+}$  and the $^3P_1 - ^1P_1$ mixing angle for the ground state, as a function of the current quark mass $m_{f'}$, at fixed value of  $m_f  = 830 $ MeV ($\equiv m_c$). All meson masses in GeV, rounded off to the nearest 5 MeV after computing the mixing angle. The (orientative) physical points are highlighted in boldface. The experimental values of the masses from the PDG are given when available.
}
\vskip1.5mm
\label{TABLE-Axial-D}
\begin{tabular}{c | c c c c c  }
\hline
$m_{f'}$ & $^3P_1$ & $^1P_1$ &   $1 P_1 $  &  $1 P_1 ^{\prime}$ & $\theta _P$  \\
\hline
{\bf 0.000} & {\bf 2.350} & {\bf 2.490} & {\bf 2.220} & {\bf 2.600} & {\bf 34.1$^{\circ}$}\\
0.400 & 2.870 & 2.905 & 2.860 & 2.915 & 24.3$^{\circ}$\\
0.500 & 2.990 & 3.020 & 2.985 & 3.025 & 20.4$^{\circ}$\\
0.600 & 3.105 & 3.135 & 3.105 & 3.135 & 14.3$^{\circ}$\\
0.700 & 3.220 & 3.250 & 3.220 & 3.250 & 10.9$^{\circ}$\\
0.725 & 3.250 & 3.275 & 3.250 & 3.275 & 8$^{\circ}$\\
0.750 & 3.280 & 3.305 & 3.280 & 3.305 & 0$^{\circ}$\\
0.800 & 3.335 & 3.360 & 3.335 & 3.360 & 0$^{\circ}$\\ 
\hline
{\bf 0.830} & {\bf 3.370} & {\bf 3.395} & {\bf 3.370} & {\bf 3.395} & {\bf 0$^{\circ}$}\\
 {\bf State} & $\chi_{c1} (1P)$ & $h_{c} (1P)$ & &  &  \\ 
 $\mathbf{m_{exp}}$ & 3.511  & 3.525 & &  &  \\ 
\hline
0.900 & 3.450 & 3.470 & 3.450 & 3.470 & 0$^{\circ}$\\
0.925 & 3.475 & 3.500 & 3.475 & 3.500 & 8.5$^{\circ}$\\
0.950 & 3.505 & 3.525 & 3.505 & 3.525 & 8.5$^{\circ}$\\
1.000 & 3.560 & 3.580 & 3.560 & 3.585 & 8.5$^{\circ}$\\
1.100 & 3.670 & 3.690 & 3.670 & 3.695 & 14.5$^{\circ}$\\
1.200 & 3.780 & 3.800 & 3.780 & 3.800 & 16.8$^{\circ}$\\
1.300 & 3.890 & 3.910 & 3.885 & 3.910 & 18.4$^{\circ}$\\
1.400 & 3.995 & 4.015 & 3.995 & 4.020 & 21.5$^{\circ}$\\
1.500 & 4.105 & 4.125 & 4.100 & 4.125 & 26.8$^{\circ}$\\
1.800 & 4.425 & 4.440 & 4.420 & 4.460 & 26.6$^{\circ}$\\
2.000 & 4.635 & 4.650 & 4.630 & 4.660 & 28.4$^{\circ}$\\
2.300 & 4.950 & 4.965 & 4.940 & 4.975 & 30.5$^{\circ}$\\
3.000 & 5.675 & 5.690 & 5.665 & 5.700 & 31.8$^{\circ}$\\
{\bf 3.900} & {\bf 6.595} & {\bf 6.610} & {\bf 6.580} & {\bf 6.620} & {\bf 33.4$^{\circ}$}\\
\hline
\end{tabular}
\end{table}
\end{center}

\begin{figure}[th]
\centering
\includegraphics[{height=8.0cm,width=8.0cm}]{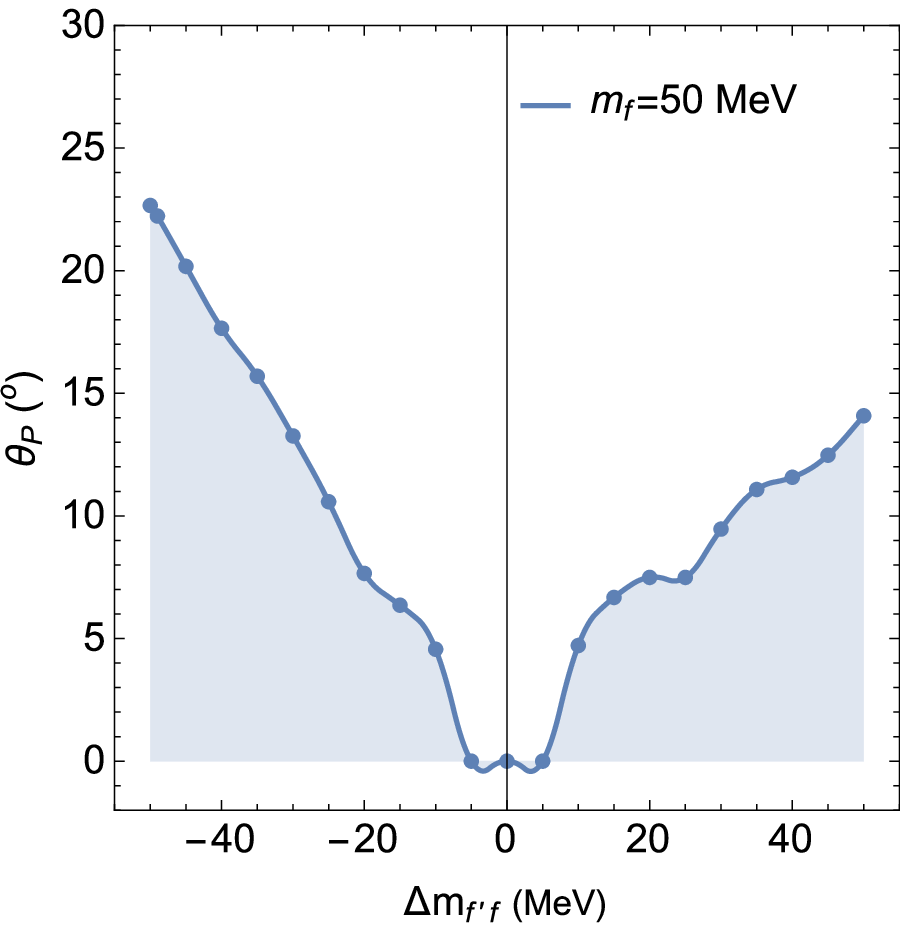} 
\includegraphics[{height=8.0cm,width=8.0cm}]{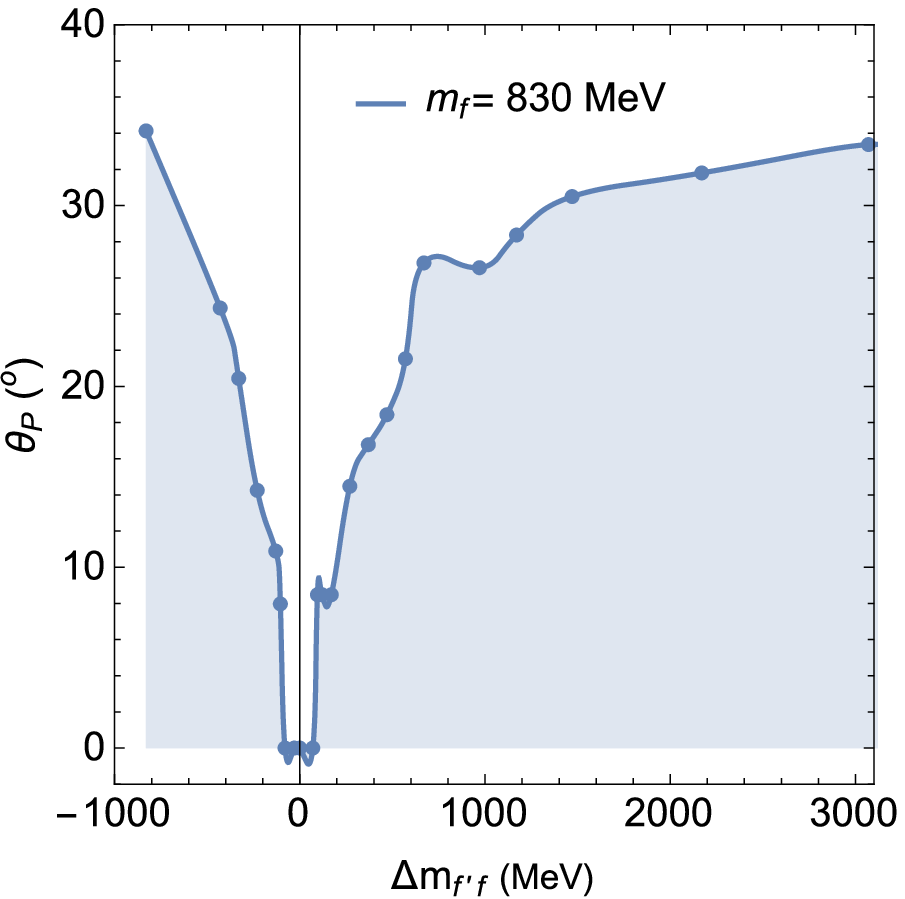} 
\caption{Mixing angle between the two $(1^{+})$ states  as a function of the current quark mass splitting $ \Delta m_{f' f} = m_{f'}-m_f $. 
Left plot: $m_f = 50 $ MeV ($\equiv m_s$). Right plot: $m_f = 830 $ MeV ($\equiv m_c$). 
The mixing angle vanishes at the deep valley in the middle of each plot (pure ${\bf L}$--${\bf S}$ coupling or $\theta_P=0$) when the quark masses are equal.
}
\label{Fig-mixing-angle-s}
\end{figure}

The mixing angle given in Tables~\ref{TABLE-Axial-S} and~\ref{TABLE-Axial-D} as function of the quark mass splitting $ \Delta m_{f' f} = m_{f'} - m_{f}$ is plotted in Fig.~\ref{Fig-mixing-angle-s}, at fixed values of  $m_f $. 

The plots show that at $\Delta m_{f' f} \approx 0$, the mixing angle also vanishes. 
As the current quark mass splitting increases, the mixing angle augments, 
tending to its maximum value  of $\theta _P = 35.3 ^{\circ}$ in the heavy quark limit of $m_f'$ for $m_f$ fixed.

The point of this exercise is not an eventual agreement or disagreement with experiment, but to clarify (given the confusion that we have seen in the literature) how the pure $q\bar{q}$ has to (roughly) behave as a function of the quark mass and how the value of the mixing angle is fixed, from theory, at the ends of the spectrum.

For completeness, we will also give two tables with numerical results for equal--flavor ($m_f=m_{f'}$),
though charge conjugation makes the mixing angle vanish. The discussion will be very brief; further outcomes of the calculation for other omitted mesons can be obtained from the authors upon request.
The first one,  Table~\ref{TABLE-Strange-Axial-PS-Vector} lists the masses of the axial vector mesons with one light quark, as function of the antiquark mass. Since we have already given the corresponding mixing angles for the cases of interest in Table~\ref{TABLE-Axial-S} above, we now compare instead the closed-flavor axial mesons with the masses of the vector and pseudoscalar mesons (with open flavor) computed with the same Coulomb approach, that eventually allows to obtain the phase space for the strong decay $1^+\to 1^- 0^-$.

\begin{center}
\begin{table}[h!]
\caption{Masses of the lowest (closed--flavor) axial, (open--flavor) pseudoscalar and vector states as a function of the current  quark mass $m_{f'} $. (The approximate strange mass within this Hamiltonian is noted.) For the pseudoscalar and vector mesons, the light quark mass is fixed at $m_f = 1$ MeV, and for the axial state $m_f = m_{f'}$, as these are the masses relevant for the decay $1^+\to K^* K$. In the pseudoscalar case, the masses are calculated within RPA approach~\cite{LlanesEstrada:2001kr,LlanesEstrada:2004wr}. The values are given in GeV.}
\vskip1.5mm
\label{TABLE-Strange-Axial-PS-Vector}
\begin{tabular}{c | c c c  }
\hline
$m_{f'}$ &  $0^+ (1^{++})$ &  $\frac{1}{2} (1^{-})$ &  $\frac{1}{2} (0^{-})$ \\
\hline
\hline
0	& 1.215	& 0.760	& 0	\\
\hline
0.005	& 1.260	& 0.770	& 0.210	 \\
\hline
0.010	& 1.295	& 0.780	& 0.290	\\
\hline
0.020	& 1.355	& 0.810	& 0.395	 \\
\hline
0.030	& 1.410	& 0.835	& 0.460	 \\
\hline
0.040	& 1.465	& 0.860	& 0.515	 \\
\hline
{\bf 0.050}	& {\bf 1.515}	& {\bf 0.890}	& {\bf 0.560}	 \\
{\bf State} & $\mathbf{f_1(1420)}$ & $\mathbf{K^*(892)}$ & $\mathbf{K(497)}$ \\
 $\mathbf{m_{exp}}$   & {\bf 1.426(1)}  & {\bf 0.892} & {\bf 0.498}  \\ 
\hline
0.060	& 1.560	& 0.910	& 0.595	\\
\hline
0.070	& 1.605	& 0.935	& 0.630	\\
\hline
0.080	& 1.650	& 0.960	& 0.665	 \\
\hline
0.090	& 1.690	& 0.980	& 0.690	 \\
\hline
0.100	& 1.735	& 1.005	& 0.720	 \\
\hline
0.110	& 1.775	& 1.025	& 0.745	\\
\hline
0.120	& 1.815	& 1.045	& 0.770 \\
\hline
\hline
\end{tabular}
\end{table}
\end{center}

Towards the end of the table, for a quark mass a bit above twice the strange mass, the axial vector state becomes bound. This is the situation empirically found for charmonium, where the ground state $1^{++}$ and $1^{+-}$ mesons are bound, and it is the $X(3872)$ or $\chi_1'(3872)$, whose $q\bar{q}$ component has one radial excitation, that finds itself at the strong decay threshold to $D^*D$.

Therefore, Table~\ref{TABLE-Strange-Axial-PS-Vector} is continued in Table~\ref{TABLE-Charm-Axial-PS-Vector} but not with the first $1^3P_1$ ground state axial--vector meson, rather with its first radial excitation $2^3P_1$ that is relevant for the charm region.

\begin{center}
\begin{table}[h!]
\caption{Masses of the (closed--flavor) axial, (open--flavor) pseudoscalar and vector states as function of the current quark mass $m_{f'}$. (The approximate charm mass in this Hamiltonian is highlighted.) For the pseudoscalar and vector mesons, these are the lowest states obtained with light quark mass fixed at $m_f = 1$ MeV; and for the axial meson this is the $2 ^3 P_1 $ state with $m_f = m_{f'}$. The values are given in GeV.}
\vskip1.5mm
\label{TABLE-Charm-Axial-PS-Vector}
\begin{tabular}{c | c c c}
\hline
$m_{f'}$ &  $2 ^3 P_1 $  &  $\frac{1}{2} (1^{-})$ &  $\frac{1}{2} (0^{-})$ \\
\hline
\hline
0.600	&  3.330	&  1.790	& 1.730  \\
\hline
0.650	&  3.440	&  1.855	& 1.800	\\
\hline
0.700	& 3.545	& 1.920	& 1.865	\\
\hline
0.750	& 3.655	& 1.985 	&  1.930	\\
\hline
0.800	&  3.760 &  2.050 &  1.995 \\
\hline
{\bf 0.830}	& {\bf 3.825}	& {\bf 2.085}	& {\bf 2.035}	\\
 {\bf State}  & $\mathbf{X(3872)}$ & $\mathbf{D^{*} }$ & $\mathbf{D }$ \\
 {\bf $\mathbf{m_{exp}}$ (charged)} & - & {\bf 2.010} & {\bf 1870} \\
 {\bf $\mathbf{m_{exp}}$ (neutral)} & {\bf 3.872}& {\bf 2.007} & {\bf 1865} \\
\hline
0.850	&  3.870 &  2.110 &  2.060 \\
\hline
0.900	& 3.975	&  2.170 & 2.125	\\
\hline
0.950	&  4.080	& 2.235	& 2.185	\\
\hline
1.000	&  4.185	& 2.295 	& 2.250	\\
\hline
1.050	&  4.290	& 2.355	& 2.310	\\
\hline
1.100	&  4.395	& 2.410	& 2.370	\\
\hline
1.150	& 4.500	& 2.470	&  2.430 \\
\hline
1.200	& 4.605 & 2.530	& 2.490	\\
\hline
\hline
\end{tabular}
\end{table}
\end{center}

Once more, the listed state, corresponding to the first radial excitation, passes from being above threshold and decaying strongly, to becoming a bound state under threshold. In the model, this happens for energies below the charmonium spectrum, but this is because the threshold comes too high (the calculation of the mass of the $D$ meson seems to be overshooting). However in nature the cross from unbound to bound happens for quark masses so close to the actual charm mass, that the state is pegged at the threshold and the discussion of how much of its nature is due to its quarkonium seed and how much to its molecular component~\cite{Kalashnikova:2010hv} has generated an inmense literature
(see for example~\cite{Barnes:2003vb,Meng:2007cx,Ferretti:2013faa}).

Once a number of numerical results has been exposed, we return to the open--flavor case, where the discussion of the axial--vector mixing angle is germane, and discuss three additional physics topics.

\newpage
\section{Some physical consequences}
In this section we explore several contemporary physical consequences and applications in meson spectroscopy.

Subsection~\ref{subsec:Bc} highlights the $B_{c1}$ axial vector mesons because a chain of reasoning based on LO Heavy Quark Effective Theory suggests that the mixing angle in the second excited state can be directly read off  from experiment, checking whether indeed the mixing is near the ideal $s_Q$--$j_q$ coupling.
\subsection{Decays of excited $B_{c1}$ mesons} \label{subsec:Bc}

The $B_c$ family of mesons is composed of one $b$--quark and one $c$--antiquark. 
Because $m_b>>m_c$, Heavy Quark Spin Symmetry dictates that the $b$--spin $s_b$ is a good 
quantum number and the $s_b$--$j_c$ coupling (\emph{aka} $j$--$j$ coupling) applies.
This is supported by the early NRQCD computation in a quenched lattice of~\cite{Davies:1996gi}
that finds a mixing angle $\theta_P=(33.4\pm 1.2)^{\rm o}$ quite near the extreme Heavy--Light value
of Eq.~(\ref{maxangle}).

No axial vector mesons with this flavor content have been experimentally reported yet. A simple linear interpolation between the $c\bar{c}$ and $b\bar{b}$ spectra, leaning on the known masses of the $\eta_c$, $\eta_c(2S)$, $B_c$, $B_c(2S)$, $\eta_b$, $\eta_b(2S)$, $\chi_c$, $h_c$, $X(3872)$, $\chi_b$, $\chi_b(2S)$ and $h_b(2S)$ suggests that the first two pairs of $B_{c1}$ axial vector mesons are to be found near $6780\pm 30$ MeV and $7130\pm 30$ MeV.

The threshold for the strong $s$--wave decay $1^+\to 1^-0^-$ of an axial meson is, for the $B_{c1}$ family, given by the two energies
$m_B+m_{D^*}=(5.279+2.010)$ GeV $= 7.289$ GeV and
$m_{B^*}+m_D=(5.325+1.870)$ GeV $= 7.195$ GeV respectively. Thus, all four of the first $B_{c1}$ mesons will be narrow bound states, just like in the $b\bar{b}$ spectrum.

The actual Coulomb gauge model calculation shown in table~\ref{tab:Bc} concurs with this observation. Though it seems likely that the computed $B_{c1}$ masses lie 100 MeV too low, it seems clear that it is the third pair of $B_{c1}$ mesons (in the Coulomb approach, around 7.31 GeV, in the real world probably up to 7.4 GeV, as supported by the recent model of~\cite{Akbar:2018hiw} and references therein) that will be able to decay strongly.
\begin{table}
\caption{\label{tab:Bc} Computed masses (in GeV) of pseudoscalar, vector, and axial--vector $B_{c}$ mesons in the Coulomb gauge model with $m_g=0.6$ GeV, $C_h=0.7$, $m_b=3.9$ GeV, $m_c=0.83$ GeV.
Also shown are the masses of the two experimentally known pseudoscalar states and a guess (based on interpolating between charmonium and bottomonium with the states listed in the text) at the mass of the $B_{c1}$. It is clear that the third pair of $B_{c1}$ mesons (highlighted in boldface), probably around {\bf 7.4} GeV in view of all the information available, will be the lightest one that can decay into the open flavor channels $BD^*$ and $B^*D$. (All calculations rounded off to the nearest 5 MeV.)}

\begin{tabular}{c||ccccc} \hline
$0^-$ ($B_c$) &          6.310 &       6.770 &       7.135       &       7.440 &       7.710 \\ 
$\mathbf{m_{exp}}$  & 6.275(0.8)&   6.871(2) &                   &             &             \\ \hline       
$1^-$ ($B_c^*$) &  6.325&   6.780    &   7.140           &  7.445      & 7.715       \\ \hline    
$^3P_1$  &     6.595    &   6.980    &   7.305     &  7.590      & 7.845       \\
$^1P_1$  &     6.610    &   6.990    &   7.315     &  7.595      &  7.850      \\ \hline
Mixed $1^+$ &     6.580, 6.620    &  6.975, 7.000     & {\bf 7.300, 7.320}        &        &             \\ 
Interpolated & $6.780\pm 0.030$ &   $7.130\pm 0.030$ &  & & \\
from $c\bar{c}$, $b\bar{b}$ & & & & & \\ \hline 
Potential model of~\cite{Akbar:2018hiw} & 6.725, 6.744& 7.098, 7.105& 7.393, 7.405 & & \\  \hline
Instantaneous BS~\cite{Li:2018eqc} & 6.815,6.830 & 7.168, 7.174 & & & \\ \hline
Lattice (quenched NRQCD)~\cite{Davies:1996gi} & 6.738(8), 6.760(8) & & & & \\
\hline \hline
\end{tabular}
\end{table}

\begin{figure}
\begin{minipage}{0.45\textwidth}
\begin{center}
\includegraphics[width=\textwidth]{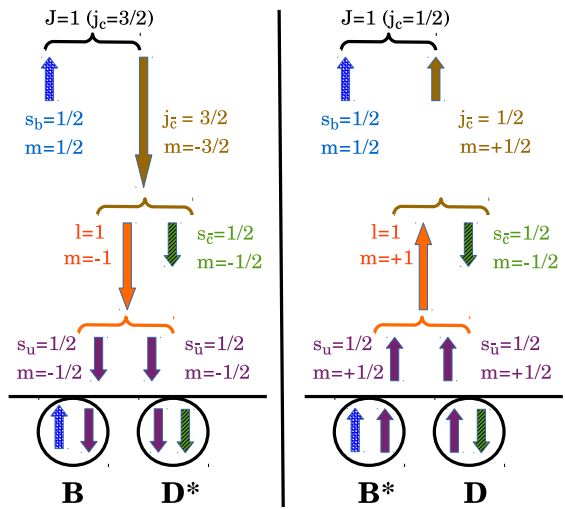}
\end{center}
\end{minipage}\ \ \ 
\begin{minipage}{0.45\textwidth}
\begin{center} \vspace{-2cm}
\caption{\label{fig:Bcdecays} The decay of the excited $B_{c1}$ axial vector meson selects
$BD^*$ or $B^*D$ (in the heavy quark limit) depending on the internal angular momentum 
$j_c=3/2$ or $j_c=1/2$. The idea is that the heavy quark spins (hatched) are unaffected. Thus, the $s_b$ $b$--quark spin goes directly into the final state. When the string snaps and a light quark--antiquark pair is created, the $s_c$ is supposedly likewise not affected, because $m_c\gg \Lambda_{\rm QCD}, m_u$. In that case, conservation of angular momentum as indicated predicts that each of the $j_c$ states decays to a different vector/pseudoscalar combination distinguishable by flavor.}
\end{center}
\end{minipage}
\vspace{-4cm}
\end{figure}

Once the spectroscopy has been reviewed, we can discuss how the mixing angle of the $1^+$ $B_c$ mesons that can decay to open flavor channels can be exposed~\footnote{There is a recent  preprint~\cite{Li:2018eqc} that provides a complementary point of view from the Dyson--Schwinger Equations.}. The guiding principle is that, in the decay process, light degrees of freedom cannot alter the spin of a heavy quark. This means that $s_b$ for the heavy quark can be read off directly in the final state. Once this has been discounted,
the interesting observation is that whether $j_c=1/2$ or $j_c=3/2$ for the charm quark can also be tracked to the final state. This is because the light degrees of freedom (eventually, a constituent $u\bar{u}$ pair) cannot flip the spin of the charm quark either. This is illustrated in figure~\ref{fig:Bcdecays}.

The outcome is that, out of the two possible channels, the configuration with $j_c=3/2$ selects $B\bar{D}^*$, and this is distinguishable from $B^*\bar{D}$ which is how the $j_c=1/2$ state decays.
Thus, the almost perfect $s_Q$--$j_q$ mixing for this excited $B_{c1}$ states can be read off the final state. Uncertainties in the prediction are $O(m_c/m_b)$ from demanding that the $s_b$ spin stays constant in the decay, and a smaller $O(m_u/m_c)$ from fixing the charm spin, amounting to a 25\% uncertainty (this is still good enough to allow clear distinction of the two channels).

It remains to hope that a future upgrade of Belle-II (or some other $B$ factory) can thoroughly explore the much unknown $B_c$ spectrum extending above 12.55 GeV. In addition to the one highlighted here, there are many physics opportunities in doing so~\cite{Drutskoy:2012gt}.

\subsection{Mixing of strange $K_1$ mesons}

The literature is riddled with discussion about the correct angle mixing the $K_1$ mesons composed by a strange and a light quark--antiquark pair with isospin 1/2. 
The $^3P_1$ $SU(3)$ nonet containing $a_1(1260)$, whose $J^{PC}=1^{++}$ eigenstates are $f_1(1285)$ and $f_1(1420)$, would contain four $K_{1A}$ kaon resonances. Likewise, the 
$^1P_1$ $SU(3)$ nonet containing $b_1(1235)$, whose $J^{PC}=1^{+-}$ eigenstates are $h_1(1170)$ and $h_1(1380)$, would contain four $K_{1B}$s.
The rotation matrix equivalent to Eq.~(\ref{jqrotation}) is
\begin{equation} \label{Krotation}
\left( 
\begin{array}{c}
$$K_1(1400)$$ \\  $$K_1(1270)$$
\end{array}
\right)= 
\left( 
\begin{array}{cc}
$$\cos \theta_{K_1}$$ & $$-\sin \theta_{K_1}$$ \\
$$\sin \theta_{K_1}$$ & $$\cos \theta_{K_1}$$ 
\end{array}
\right)
\left( 
\begin{array}{c}
$$K_{1A}$$ \\  $$K_{1B}$$
\end{array}
\right)\ .
\end{equation}
Typical analysis examines the masses, with formulae such as
\begin{eqnarray} \label{simplemasses}
m^2_{K_{1A}} = m^2_{1400}\cos^2\theta_{K_1} + m^2_{1270}\sin^2\theta_{K_1}, \\ \nonumber
m^2_{K_{1B}} = m^2_{1400}\sin^2\theta_{K_1} + m^2_{1270}\cos^2\theta_{K_1}.
\end{eqnarray}
For example, Suzuki~\cite{Suzuki:1993yc} proposed that $\theta_{K_1}=33^{\rm o}$ or alternatively $57^{\rm o}$. Cheng finds a smaller value of order $(28-30)^{\rm o}$~\cite{Cheng:2013cwa} excluding the larger solution. Isgur and Godfrey seem to quote some $34^{\rm o}$~\cite{Godfrey:1985xj} 
($56^{\rm o}$ taking the complementary angle).
Burakovsky and Goldman quote a large mixing between 35 and 55 degrees~\cite{Burakovsky:1997dd}
and another nonrelativistic quark model by Li and Li~\cite{Li:2006we} yields 59 degrees (the complementary angle of 31 degrees, of course).

To isolate the rotation angle, one needs the masses on the left hand side of Eq.~(\ref{simplemasses}) that correspond to no physical particle; they are sometimes obtained from a model Hamiltonian, as is our case, or else they can be isolated from flavor analysis in the Gell-Mann-Okubo spirit. In this case, the difficulty is that the $f_1$ and $h_1$ mesons, because there are two in each multiplet, undergo singlet--octet flavor mixing, and that flavor angle\cite{Yang:2010ah,Dudek:2011tt,Liu:2014doa} becomes entangled with the spin angle of interest for Eq.~(\ref{Krotation}).

In any case, we do not concur here with the findings in the literature. Our result for the mixing angle can be read off the left end of the left plot in figure~\ref{Fig-mixing-angle-s} and has been highlighted in bold face in table~\ref{TABLE-Axial-S}, and it is about 22.2$^{\rm o}$, well below the $O(30^{\rm o})$ coming from phenomenological analysis.
First, it is easy to identify the difference: we use model masses computed within the same Hamiltonian, instead of employing phenomenological masses read off from the experiment. This would seem like a shortcoming on our part. But let the reader consider that the asymptotic value for the mixing angle, 
35.2$^{\rm o}$ from Eq.~(\ref{maxangle}), should not be reached as early as a light--strange system: we do not theoretically expect angles of order 30 degrees until the light--charm or light--bottom mesons.

That is, our calculation is actually closer to theory expectations in this regard. The extractions from the experimental masses are probably parametrizing other more complex physics (such as meson--molecule mixing, decay channel influence, or mixing with further mesons) into this pure $q\bar{q}$ mixing angle, where it does not belong.

\subsection{Excited mesons with light quarks and insensitivity to chiral symmetry breaking}
\label{subsec:excited}

As is well known, the QCD Lagrangian admits an approximate chiral symmetry
due to the extreme lightness of the up and down quarks, $\left( m_u,\ m_d \right) \sim O(1-10{\rm MeV}) \ll
\left(4\pi f_\pi,\ m_N \right) \sim O(1{\rm GeV})$.
In terms of the running quark mass $m(k)$, the chiral charge associated to this approximate global symmetry is given as~\cite{Bicudo:2009cr}

\begin{eqnarray} \label{chiralcharge}
Q_5^a  = \int \frac{d^3k}{(2\pi)^3} 
\sum_{\lambda \lambda ' c}\ \ \sum_{\{f f'\}_{\rm light }} \left(  \frac{\tau^a}{2} \right)_{ff'}
{ k \over \sqrt{ k^2 + m^2(k)}}
 \\ \nonumber 
\left( ({\boldsymbol{\sigma}}\cdot {\bf
\hat{k}})_{\lambda \lambda'}  	
\left( B^\dagger_{k \lambda f c} B_{k \lambda' f' c} + D^\dagger_{-k \lambda' f' c}
D_{-k \lambda f c}
\right) + \right. \\  \nonumber \left.
{ m(k) \over k} (i\sigma_2)_{\lambda \lambda'} \
\left( B^{\dagger}_{k\lambda f c} D^\dagger_{-k\lambda'f'c}+
B_{k \lambda' f' c} D_{-k \lambda f c}
\right) \right) \ .
\end{eqnarray}

Spontaneous chiral symmetry is triggered by the apparition of a nonnegligible $m(k)$ (recall figure~\ref{fig:massgap}). Then, for low $k$, the term  in the last line dominates. This creates or destroys the $q\bar{q}$ Fock--space component of a pion (which implements the nonlinear, Goldstone--boson realization of chiral symmetry)~\footnote{The corresponding sine of the gap angle $\sin\phi(k) = \frac{m(k)}{\sqrt{m^2+k^2}}$ is actually the pion wavefunction in the Random Phase Approximation.}.

However, if the typical $k$ is large, because quarks are in an excited state, then $m(k)/k$ can become small. In consequence, that third line gets to be negligible and the second line turns dominant, noticing that it counts the number of light quarks and antiquarks but applies a $({\boldsymbol{\sigma}}\cdot {\bf
\hat{k}})$ operator to them. Thus, hadron states in which all light quarks sit in wavefunctions with large momentum compared to $m(0)$, the constituent quark mass, are expected to come in mass--degenerate multiplets.  

As is easily seen, $[Q_5^a,P]\neq 0$ (in the Pauli--Dirac representation, $P=\int \bar{\Psi} \gamma^0 \Psi$, for example), so that the chiral charge cannot be diagonalized simultaneously with parity  though both commute with the Hamiltonian.  Because parity is an exact symmetry of the strong interactions and thus conserved by them, it is normally chosen as an observable. The action of the chiral charge then generates chiral multiplets that contain members of opposite parities~\cite{Detar:1988kn}. (If the quark and antiquark flavors are equal, then charge conjugation $C$ is also conserved, but $[Q_5,C]\neq 0$, so the same comment applies.) 
As befits this work on $J=1$ mixing, we will dedicate some of the discussion in the next subsection~\ref{subsec:J1doubling} to the angular--momentum structure of the chiral charge.

For now, let us just remark that a prediction of QCD as a chiral theory could be that there is a parity doubling (chiral symmetry is closer to Wigner than to Goldstone mode in the high--energy spectrum). 
This insensitivity of the high spectrum to chiral symmetry breaking has been estimated to trigger around 2.5 GeV in the light meson spectrum~\cite{Swanson:2003ec}, a challenging but not outlandish scale.

There are two levels of discussion. 
The first is whether the parity doubling happens at all in the experimental spectrum (it does in
adequate models of QCD that can address high excitation; model independent approaches such as lattice gauge theory and effective theories have difficulties in credibly doing so, and there are not many solid statements, but see~\cite{Denissenya:2014poa} and references therein for a briefing).
This is not an idle question. At the present time, the experimental evidence is marginal (Regge trajectories do not clearly converge, and the assignment of various parity doublers in the spectrum is debatable). The reason one can theoretically question the very interesting concept is because the Wigner realization of the symmetry requires $k\gg m(k)$. If an excited meson is mostly $q\bar{q}$, then
$\langle k \rangle \sim \frac{M_{\rm meson}}{2} \gg m(\langle k\rangle)$, and we expect parity doubling to set in. However, for multiquark mesons, $\langle k \rangle \sim \frac{M_{\rm meson}}{N} $ can well be of the same order of $m(k)$; then, quark velocity is small and chiral symmetry continues in Goldstone mode even for very excited mesons.

The second level arises when and if the parity doubling is well established. In that case, one could ask how fast does the high spectrum become insensitive to chiral symmetry breaking. For example,
a constituent quark model with explicit (not spontaneous) chiral symmetry breaking would have $m(k)=m_q$ a constant. Then one would expect~\cite{Segovia:2008zza}, as with any relativistic corrections (since, as seen in Eq.~(\ref{chiralcharge}), the action of the chiral charge is to hit the quark spin with the operator $\boldsymbol{\sigma}\cdot {\bf k}$), to have 
\begin{equation}
(M_+-M_-)\propto \frac{1}{M_++M_-},
\end{equation}
that is, the splitting would fall--off as the inverse of the state mass, $\Delta\propto 1/M$.
However, in QCD and models thereof that implement spontaneous chiral symmetry breaking, $m(k)$ falls with $k$, so that the parity doubling should happen faster, depending on the support of the mass gap function and of the quark wavefunction inside the hadron. Finding parity degeneracy in the spectrum
can help learning about the quark mass function~\cite{Bicudo:2009cr}. 
From all the information available, this might be possible for high--$J$ excitations, but it seems too unlikely for fixed--$J$, large radial $n_r$ excitations, as we will next reconfirm.

\subsection{Excited $J=1$ mesons and parity doubling} \label{subsec:J1doubling}

Let us focuse on the angular momentum part of the chiral charge and abstract all other features (flavor, color, radial parts of the wavefunction). The content of its action with respect to angular momentum in this high--$k$ regime is in the $({\boldsymbol{\sigma}}\cdot {\bf
\hat{k}})_{\lambda \lambda'} $ scalar product. This is a $^3P_0$ structure with $L=1$ coupled to $S=1$ to yield $J=0$. Therefore, $1^+$ states are mapped to $1^-$ states, to satisfy the parity flip and the angular momentum addition rules with a scalar operator.

In a state with several light quarks/antiquarks of different momentum, most straightforwardly in a baryon~\cite{Bicudo:2009cr}, one can apply $({\boldsymbol{\sigma}}\cdot {\bf \hat{k}_1})({\boldsymbol{\sigma}}\cdot {\bf \hat{k}_2})\dots$ more than once. But in $q\bar{q}$ mesons, because there is only one $k$ and $({\boldsymbol{\sigma}}\cdot {\bf \hat{k}})^2  = \boldsymbol 1$, we can speak of parity doublets (modulo isospin structure).

There are two conceivable ways of achieving large $k$. Probably the clearest one is to choose increasingly large $J$~\cite{Bicudo:2009cr}. But at fixed $J=1$ as relevant for this paper, one has to examine highly (radially--) excited states.

We have carried out a computation of several radial excitations for axial--vector mesons and compared it to one of vector mesons in table~\ref{tab:quiral}. We have stopped reporting further excitations upon reaching the charmonium region, since numerous unrelated resonances start appearing there and clutter the spectrum.

\begin{table}
\caption{\label{tab:quiral} Numeric computations of the vector and axial vector mesons (in MeV) with equal light flavor $m_f=m_{f'}=1$ MeV up to the fifth excitation, to examine the concept of insensitivity to chiral symmetry breaking in the high spectrum. The evidence for this insensitivity is marginal, with the splitting between would--be chiral partners falling as $M^{-1.2}$, only slightly faster than the natural $M^{-1}$ fall--off for relativistic interactions in constituent quark models~\cite{Segovia:2008zza}. }
\begin{tabular}{c|ccccc}\hline
$1^{++}$  & 1035 & 1740 & 2305 & 2780 & 3190 \\
  State  & $a_1(1260)$ & $a_1(1640) $ & & & \\
  $m_{exp}$  &   1230(40)  & 1640(40) & & & \\
\hline
$1^{+-}$  & 1270 & 1870 & 2400 & 2850 & 3245 \\
  State & $ b_1(1235) $ &  & & & \\
  $m_{exp}$  &   1230(3)  &  & & & \\
\hline
$1^{--}$  &  730 & 1515 & 2115 & 2610 & 3040 \\
  State  & $ \rho(770) $ & $ \rho(1450) $ & $ \rho(1700) $& & \\
  $m_{exp}$  &   775(0.3)  & 1465(25) & 1720(20) & & \\
\hline
$1^{--'}$ & 1320 & 1955 & 2485 & 2935 & 3330 \\ \hline
$\frac{M_{1^{++}}+M_{1^{+-}}}{2}-\frac{M_{1^{--}}+M_{1^{--'}}}{2}$ &
130 & 70 & 50 & 45 & 35 \\ \hline
\end{tabular}
\end{table}
If we just look at the actual masses, the idea of insensitivity to chiral symmetry breaking seems reasonable. But to distinguish whether this breaking is explicit (as in a constituent quark model) or spontaneous (as in QCD, the truncated Dyson-Schwinger equations thereof, or the Coulomb gauge approach here exposed) is much more difficult. The (configuration--averaged) parity splitting in the last line of the table is clearly falling.

\begin{figure}
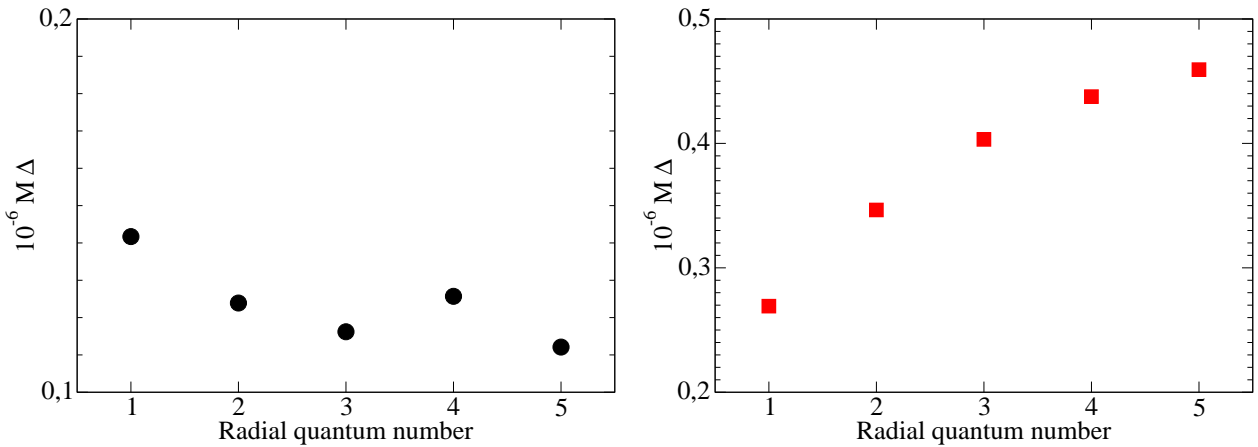

\begin{center}
\includegraphics*[width=0.45\textwidth]{FIGS.DIR/Restoration1.eps}\ \ \
\includegraphics*[width=0.45\textwidth]{FIGS.DIR/Restoration2.eps}
\caption{\label{fig:insensitivity} Left: We plot $10^{-6}M \Delta$ with $M=\frac{1}{4}\sum_{i} M_i ^{(J=1)} $ ($M$ and $\Delta$ given in units of MeV) for each radial excitation (multiplet average mass) and $\Delta= \bar{M}_{1^+}-\bar{M}_{1^-}$ is the parity splitting between the average of the two axial--vector masses and that of the two vector masses for each radial quantum number $n$. 
This observable would be about flat for a constituent quark model, since $k\gg m_q$ would damp the angular--momentum (and hence parity) splittings as $1/M$ (which we are correcting for). That it falls with radial quantum number is a feeble indication of insensitivity to spontaneous chiral symmetry breaking in the upper spectrum.\\
Right: A related observable constructed from the computations of Glozman and Wagenbrunn~\cite{Wagenbrunn:2006cs}, $10^{-6}(M_{1^{++}}-M_{1^{--}})\frac{M_{1^{++}}+M_{1^{--}}}{2}$, with the vector meson mixing chosen to be in the same representation of the chiral group as the $1^{++}$ meson. In this case, the onset of insensitivity to the quark mass is even slower than the $1/M$ fall--off expected in a constituent quark model. }
\end{center}
\end{figure}

For this purpose we have constructed the quantity $( \overline{M}_{1^+}-\overline{M}_{1^-})\frac{1}{4}\sum_{i} M_i ^{(J=1)}  $ or, for short, $M\Delta$, and plot it in the left panel of figure~\ref{fig:insensitivity}. It is clear that this quantity is proportional to the parity splitting $\Delta$ but correcting its $M$ dependence: in the quark model, it should flatten out as the cost of one unit of $L$ is down by one power of $M$.

The data in the figure shows that the chiral model here discussed falls slightly faster, but not by much. A fit to the computer data yields $M^{-\alpha}$ with $\alpha\simeq 1.2$ just above 1. Thus, we conclude that gaining information about the running quark mass from the radial--like excitations of the $J=1$ mesons is not to be realistically expected, unlike perhaps the large-$J$ excitations.

Now, the calculation that we report is almost identical to the one in~\cite{Wagenbrunn:2006cs} for this particular channel (we are using slightly different parametrizations of the Coulomb--like potential but the setup is very similar). They quote their states in terms of the string tension $\sqrt{\sigma}$, but once pinned by making the $\rho$ mass in both calculations equal, the differences are at most 50 MeV and this only for quite excited states. Unsurprisingly, their results in what concerns insensitivity to chiral symmetry breaking are very similar.

Since they address vector--meson $s-d$ wave mixing carefully, they can identify the chiral partner of each of the axial vector mesons for equal quark flavor. The parity doublings for the $1^{+-}$ state and its $1^{--}$ partner are not clearly decreasing with $M$, but those for $1^{++}$ and corresponding $1^{--}$ in the same representation of the chiral group are indeed falling. For this last case we construct a similar quantity to our $M\Delta$, namely $10^{-6}(M_{1^{++}}-M_{1^{--}})\frac{M_{1^{++}}+M_{1^{--}}}{2}$ that only considers this doublet, and plot it in the right panel of figure~\ref{fig:insensitivity}.

This $M$-multiplied splitting is not falling: so the computation of~\cite{Wagenbrunn:2006cs} is not exposing the running quark mass $m(k)$ in this channel. This reinforces our conclusion that only the first of the two statements at the end of subsection~\ref{subsec:excited} can be addressed with $J=1$ mesons.

Finally, let us devote some discussion to the open--flavor case. Here, there is an exhaustive work~
\cite{Bicudo:2015mjf} that addresses heavy--light systems and, up to very large excitation, finds relatively slow return of the Wigner realization of chiral symmetry. 
We will only add one point related to $1^+$ meson mixing: whereas the chiral charge, due to the $\boldsymbol{\sigma}\cdot \hat{\bf k}$ operator is not diagonal in the $L$--$S$ basis, it is so in the 
$j_q$--$s_Q$ one. This comes about because
\begin{equation}\label{chiralsplit}
\left( \boldsymbol{\sigma}\cdot \hat{\bf k}\right)_{\lambda m_{sq}} = -\sqrt{4\pi} \sum_{m_sm_l} Y_1^{m_l}(\hat{\bf k}) 
\left( \boldsymbol{\sigma}^{m_s}\right)_{\lambda m_{sq}}\langle 1 m_s 1 m_l\arrowvert 00\rangle
\end{equation}
is a scalar. Since it does not act on the heavy quark, adding this total $0$ angular momentum to the light quark's $j_q$ again yields $j_q$.

By constructing the angular--momentum state of the quark (color and flavor indices are omitted)
\begin{equation}
\arrowvert j_q m_q\rangle = \int \frac{d^3k}{(2\pi)^3} R(\arrowvert {\bf k}\arrowvert)
\sum_{m_{sq}m_L} \langle \frac{1}{2} m_{sq} L m_L \arrowvert j_q m_q \rangle 
Y_L^{m_L}(\hat{\bf k}) B^\dagger_{m_{sq}} ({\bf k}) \arrowvert 0 \rangle
\end{equation}
and employing 
\begin{equation}
\left(\sigma^{m_S} \right)_{\lambda m_{sq}} = \sqrt{3} (-1)^{1-2m_{sq}}
\langle \frac{1}{2} m_{sq} 1 m_S \arrowvert \frac{1}{2} \lambda \rangle ,
\end{equation}
it is not too hard to find a closed expression for the matrix element of $Q_5$ in the $j_q$ basis connecting specific vector and axial--vector mesons, which becomes an angular momentum recoupling problem solvable by a Wigner 9$j$ symbol,
\begin{eqnarray}
\langle jq m_q \arrowvert Q_5 \arrowvert j_q m_q \rangle = \left[\int_0^\infty \frac{k^2dk}{(2\pi)^3} 
R_{1^+}^*(\arrowvert {\bf k} \arrowvert)R_{1^-}(\arrowvert {\bf k} \arrowvert) \right]
\sum_{L\Lambda}
\left(-3\sqrt{2(2L+1)(2j_q+1)}\right)
\left\{
\begin{tabular}{ccc}
$\frac{1}{2}$ & $L$ & $j_q$ \\
1 & 1 & 0\\
$\frac{1}{2}$ & $\Lambda$ & $j_q$ \\
\end{tabular}
\right\} \ .
\end{eqnarray}
The first line of the 9$j$ symbol constructs the ket from a spin 1/2 quark and the orbital angular momentum $L$. Likewise, the last line corresponds to the bra with angular momentum $\Lambda$ (in the case at hand, the values that they can take are $L=1$ and $\Lambda=0,2$ respectively). The middle line corresponds to the structure of the chiral charge in Eq.~(\ref{chiralsplit}). By columns, the first is the quark spin, the second the orbital angular momentum, and the third the total angular momentum.

This calculation shows that there is a case in which experimental data can be directly used to read off the parity splitting: for excited $B^*$--like mesons, the nearest vector and axial--vector mesons are directly chiral partners, (for other cases, they have to be disentangled from the physically mixed states).

\section{Conclusions}
\label{Conclusions}

In this work we have provided minimal background about the theoretical point of view on the problem of
axial--vector meson mixing. We have employed the Coulomb gauge Hamiltonian approach to QCD. The truncation thereof that we employ is not, at the present time, amenable to systematic improvement, nor there exists a rigorous analysis of its uncertainty. Therefore we have not fine tuned the few Hamiltonian parameters (scale of the potential $m_g$, relative strength of the transverse gluon exchange $C_h$, and quark masses at a high scale) to optimize the fit to the spectrum, though a very good fit does not seem to be within reach. Instead, the model can be used for a unified discussion of the spectrum through the whole range of quark masses.

This has allowed us to simultaneously address axial--vector $B_{c1}$ mesons, the least known of the meson systems, that share properties of heavy--light and of quarkonium systems; with kaon $K_1$ mesons, where we can qualitatively estimate the mixing angle of the $q\bar{q}$ mesons with a theoretically anchored computation and weigh on its discussion; and to address the concept of parity doubling due to the Wigner realization of chiral symmetry in the high spectrum; all within the same model and with the same interactions.

One obvious improvement that can be deployed in future work if there would be interest 
is to extend the simple one--angle analysis to a multidimensional space where the various radial 
excitations are connected by an overlap matrix. In principle, we can obtain those overlaps by integrating
the radial wavefunctions obtained $\int R^*_1(\arrowvert {\bf k}\arrowvert) R_3(\arrowvert {\bf k}\arrowvert)$ between the spin--singlet and triplet configurations. This would allow for an independent calculation of the mixing angle in taking only the ground state, but also for a refinement including the all--to--all mixing of figure~\ref{fig:mixingtype}.

But the most significant piece of work ahead is to connect the quark--antiquark formulation here presented for axial--vector mesons to one including multiquark configurations or, more directly, meson--meson ones, to be able to describe the effect of coupled channels. There is a large literature on this, especially in view of 
the effect of axial resonances in controlling the scattering of $D$ mesons in the hadron medium into which the quark--gluon plasma cools~\cite{Abreu:2011ic} or, saliently for spectroscopy,
the near--threshold $X(3872)$ meson. 
This ``cryptoexotic'' (hidden exotic) is likely~\cite{Kalashnikova:2010hv} seeded by the $q\bar{q}$ meson that we have been discussing, attracted by the threshold through mixing with molecular configurations.  At present we are considering how to address the problem with minimum effort; 
the following two observations are relevant to decide it. The first, and positive one, is that the Coulomb gauge model is a field theory. This means that higher Fock space configurations can be treated with the same parameters and on the same footing, by just extending the Fock space in which the model Hamiltonian is diagonalized. For example, meson-tetraquark mixing has been explored within this approach by other authors~\cite{Wang:2008mw}. The second, and negative one, is that the two meson--like configurations coupling to the ordinary $q\bar{q}$ will have significant momentum and boosting the 
model wavefunctions is far from trivial. Therefore, only near threshold states such as $X(3872)$ can be addressed with some confidence. Instead, for a global analysis such as we have performed here,
we would rather rely on a combination of effective theory in the spirit of~\cite{Roca:2003uk,Roca:2005nm}
where the Hamiltonian model is only used to obtain certain key coefficients at a safe kinematic point,
and the EFT is used to extend the work to arbitrary kinematics.
Finally, a competent lattice calculation~\cite{Woss:2019hse} has already been presented, so one should carefully consider in which direction can model computations complement it given their systematic limitations.

\begin{acknowledgments}

L.M.A. and A.G.F. thank the Departamento de F\'{i}sica Te\'{o}rica of the Universidad Complutense in Madrid for kind hospitality while part of this work was completed.
 Work supported by Brazilian funding agencies CNPq (contracts 308088/2017-4 and 400546/2016-7) and FAPESB (contract INT0007/2016). Additionally, MINECO:FPA2016-75654-C2-1-P (Spain); Universidad Complutense de Madrid under research group 910309 and the IPARCOS institute; and the EU's Horizon 2020 programme, grant 824093.
\end{acknowledgments}

\appendix
\section{Simplified hidden--flavor limit (quarkonium)} \label{sec:equalflavor}

In this appendix we present the computation of the axial vector meson, for the restricted case of identical flavor, as a cross--check of the earlier, more general calculation. Also, to guarantee independence of the results, we employ a slightly different formalism. Instead of the angular momentum algebra based on Clebsch-Gordan coefficients, and given that the angular momenta involved are small, we construct the wavefunctions easily employing Pauli $\sigma$ matrices multiplied by one power of orbital momentum $\hat{k}^l$ to get the p--wave. Thus,  in lieu of Eq.~(\ref{wave_func}) we have
\begin{eqnarray}
 \ket{{}^1P_1}= {\rm Const}  \frac{\delta^{mn}_{\rm color}}{\sqrt{3}} \hat{k}^l (i\sigma^2)  R(k)
\\ \nonumber 
 \ket{{}^3P_1}=i {\rm Const} \frac{\delta^{mn}_{\rm color}}{\sqrt{3}}(i\sigma^2)\epsilon_{ijl}\sigma^j\hat{k}^l R(k)\ .
\end{eqnarray}
(The notation is a bit more schematic than in Eq.~(\ref{wave_func}) but equivalent: $(i\sigma^2)$ implements $C$--conjugation, since we are coupling a particle and an antiparticle, and $R(k)$ is a shorthand for the radial wavefunction $\Psi^{nJP}_{LS}$).
As the spinor products necessary for either the longitudinal or transverse potential
are naturally expressed in terms of Pauli matrices too,
\begin{eqnarray}
U^\dagger_{\bk,s} U_{\bq,d} =\frac{1}{2}\Mask\Menosq\delta_{sd} +\frac{1}{2}\Menosk\Masq(\gk\gq)_{sd}\\ \nonumber
V^\dagger_{-\bq,s} V_{-\bk,d} = \frac{1}{2}\sigma^2\Menosk\Masq\sigma^2\delta_{sd}+\frac{1}{2}\Mask\Menosq(\sigma^2\gq\gk\sigma^2)_{sd} \\ \nonumber
U^\dagger_{\bk,s}\boldsymbol{\alpha} U_{\bq,d}=\frac{1}{2}\Mask\Menosq(\boldsymbol{\sigma}\gq)_{sd}+\frac{1}{2}\Menosk\Masq(\gk\boldsymbol{\sigma})_{sd} \\ \nonumber
V^\dagger_{-\bq,s} \boldsymbol{\alpha}V_{-\bk,d} = \frac{-1}{2}\Menosk\Masq(\sigma^2\gk\boldsymbol{\sigma}\sigma^2)_{sd}+\frac{-1}{2}\Mask\Menosq(\sigma^2\boldsymbol{\sigma}\gq\sigma^2)_{sd}\ ,
\end{eqnarray}
the computation of the TDA kernels can be carried out by taking the trace of several combinations of the $\sigma$ matrices, which is conveniently done by a symbolic calculation environment such as {\tt FORM}\cite{Ruijl:2017dtg}.

Here is how the computation of  the matrix element of the transverse potential
  \begin{align}
       \bra{{}^1P_1}H_{T}\ket{{}^1P_1}=C\sum_l \delta_{ll'}\int \dok \int \doq \ \hat{k}^l\hat{q}^{l'}\\ U(|\boldsymbol{k}-\boldsymbol{q}|)\left(\delta^{mn}-\frac{(\boldsymbol{k}-\boldsymbol{q})^m(\boldsymbol{k}-\boldsymbol{q})^n} {|\boldsymbol{k}-\boldsymbol{q}|^2}\right) T^{mn}
 \end{align}
comes out. There,
 \begin{align}
    T^{mn}= {\rm Tr}\{\sigma^2U_k^\dagger \alpha^m U_q \sigma^2 V_{-q}^{\dagger} \alpha^n V_{-k}\}
 \end{align}
that can conveniently be divided in several pieces
  \begin{align}
      & T^{mn}= T_1^{mn}+T_2^{mn}+T_3^{mn}+T_4^{mn},\\
      & T_1^{mn}=c_k c_q {\rm Tr} \{\sigma^m \gq \gk \sigma^n\},\\
      & T_2^{mn}=(1+s_k)(1-s_q){\rm Tr}\{\sigma^m \gq\sigma^n\gq\},\\
      & T_3^{mn}=(1-s_k)(1+s_q){\rm Tr}\{\gk \sigma^m \gk \sigma^n\}=\\&=(1-s_k)(1+s_q){\rm Tr}\{\sigma^m \gk \sigma^n \gk\},\\
      & T_4^{mn}=c_k c_q {\rm Tr}\{\gk \sigma^m \sigma^n\gq\}\ .
 \end{align}
The contribution proportional to the cosine of the BCS angle comes out as
 \begin{align}
   &x\left(\delta^{mn}-\frac{(\boldsymbol{k}-\boldsymbol{q})^m(\boldsymbol{k}-\boldsymbol{q})^n} {|\boldsymbol{k}-\boldsymbol{q}|^2}\right) (T_1+T_4)^{mn}=\\&=c_k c_q \Big\{3x^2-\frac{x^2(k^2+q^2)-2x^3kq}{|\boldsymbol{k}-\boldsymbol{q}|^2}\Big\}=\nonumber\\&=c_k c_q \Big\{3x^2-\frac{x^2(k^2+q^2-2xkq)}{|\boldsymbol{k}-\boldsymbol{q}|^2}\Big\}=c_kc_q2x^2\ ,
 \end{align} 
and that for the sine reads
\begin{align}
 & x\left(\delta^{mn}-\frac{(\boldsymbol{k}-\boldsymbol{q})^m(\boldsymbol{k}-\boldsymbol{q})^n} {|\boldsymbol{k}-\boldsymbol{q}|^2}\right)T_2^{mn}=\nonumber\\&=(1+s_k)(1-s_q)\Big\{-\frac{x}{2}-\frac{\frac{x}{2}(q^2-k^2)-x^2kq+x^3k^2}{|\boldsymbol{k}-\boldsymbol{q}|^2}\Big\}\ .
\end{align}
These, and the equivalent expressions for the longitudinal potential, can be expressed in terms of the $V_n$, $U_n$ angular integrals of the potentials, and the auxiliary functions derived thereof, $W_n$, $Z_n$, from Eqs.~(\ref{W_func}) and~(\ref{Z_func}).
With these, the $^1P_1$ meson reads
\begin{align}\label{kernel1}
    K^{+-,+-}&= 2 c_k c_q (V_2- 2 W_1) + (1 - s_k s_q) 4 U_1 + (1+s_ks_q) 2 V_1
\end{align} 
and that for the $^3P_1$ meson in turn
 \begin{align} \label{kernel3}
      &K^{++,++}= c_k c_q ( V_0+V_2+U_0+U_2- 2 W_1) +
     (1+s_ks_q) 2 V_1+ (1-s_ks_q) (k^2+q^2) Z_1 +  (s_k-s_q) (k^2-q^2) Z_1 \ .
 \end{align}

Because the flavors of the quark and the antiquark are the same, the computation of the eventual mixing matrix elements
$\bra{{}^3P_1}H_{L}\ket{{}^1P_1}$ and $\bra{{}^3P_1}H_{T}\ket{{}^1P_1}$ directly leads to zero. (This happens as all the traces over $\sigma$ matrices involves an odd number of them.)

We have coded the kernels in Eqs.~(\ref{kernel1}) and~(\ref{kernel3}) and used them with the computer programme of~\cite{LlanesEstrada:2004wr} for the vector mesons (that couple $s$ and $d$--waves as explained in that reference). 

First, a quick run with simple quark masses yields the vector meson masses in Table~\ref{1--}. The numeric values therein suggest that the charm and bottom masses have to be taken slightly smaller (as we have done in the main body in the paper, at $0.83$ and $3.9$ GeV respectively).

\begin{table}
\caption{Masses of $1^{--}$ closed--flavor quarkonium mesons in GeV.\label{1--}}
\begin{tabular}{|lll|lll|}
\hline
$m_q$ & $m_{1^{--}}$ & meson, $m_{exp}$\\
\hline \hline
$10^{-3}$ & $0.75$ & $\rho,0.77$\\ \hline
$5\times10^{-3}$&$0.79$&$\omega,0.78$\\ \hline
$10^{-1}$&$0.92$&$\phi,1.02$\\ \hline
$1$&$3.23$&$J/\psi,3.10$\\ \hline
$4$ &$ 9.41$&$\gamma,9.46$\\ \hline
\end{tabular}
\end{table}

Independently of that, we have then run the same code for the axial vector mesons (which seem to be requiring a quark mass slightly higher: their absolute splitting from the vector mesons does not appear to be very well captured by the model Hamiltonian).
The exercise is reported in Tables~\ref{1+-} and~\ref{1++}. 
\begin{table}
\caption{Masses of $1^{+-}$ closed--flavor quarkonium mesons in GeV. \label{1+-}}
\begin{tabular}{|lll|lll|}
\hline
$m_q$ & $m_{1^{+-}}$ & meson, $m_{exp}$\\
\hline \hline
$10^{-3}$ & $1.30$ & $h_1,1.17$\\ \hline
$5\times10^{-3}$&$0.79$&$b_1,1.23$\\ \hline
$10^{-1}$&$1.43$& \\ \hline
$1$&$3.65$&$h_c,3.52$\\ \hline
$4$ &$ 9.90$&$h_b,9.90$\\ \hline
\end{tabular}
\end{table}

\begin{table}
\caption{Masses of $1^{++}$ closed--flavor quarkonium mesons in GeV.\label{1++}}
\begin{tabular}{|lll|lll|}
\hline
$m_q$ & $m_{1^{++}}$ & meson, $m_{exp}$\\
\hline \hline
$10^{-3}$ & $1.07$ & $a_1,1.26$\\ \hline
$5\times10^{-3}$&$1.23$&$f_1,1.28$\\ \hline
$10^{-1}$&$1.21$& \\ \hline
$1$&$3.59$&$\xi_{c1},3.51$\\ \hline
$4$ &$ 9.89$&$h_{b1},9.89$\\ \hline
\end{tabular}
\end{table}
We can see that the $1^{+-}$ states are systematically higher than the $1^{++}$ ones.

To see such ordering of the two $1^+$ states analytically, let us take the simple limit $M_q \rightarrow 0$ in the TDA equation (that is, we decouple the two body and the one--body problems, which is poor field theory so we refrain from quoting any masses), in which case  $\sin(\phi(k))=s_k\rightarrow 0, \ c_k\rightarrow 1$ which simplifies the algebra very much. Additionally, we set to zero the transverse potential by
$ C_h\rightarrow 0 $, leaving only the longitudinal one $V$;
In that case, the TDA kernels become:
\begin{align}
    & K^{+-,+-}\rightarrow V_1+V_2 , \\
    & K^{++,++}\rightarrow \frac{1}{2}V_0+V_1+\frac{1}{2}V_2 \ .
\end{align}
so that $K^{+-,+-}-K^{++,++}\rightarrow \frac{V_2-V_0}{2} > 0$.
This quantity is positive because $V$ is strictly negative, and weighing its angular integrals as in Eq.~(\ref{ang_int}), the power of $x^2<1$ in $V_2$ makes it smaller than $V_0$ (which is multiplied by 1 for the polar integral). $(V_2-V_0)/2$ is not a very large quantity (because the potential $V$ is larger at $x\sim 1$  where $x^2$ is not so different from $1$ itself) but it is definitely positive. 
Thus, in this toy limit, the $1^{+-}$ meson has higher mass,
\begin{equation}
M_{1^{+-}}>M_{1^{++}}\ .
\end{equation}
This is exactly what Tables~\ref{1+-} and~\ref{1++}, are showing. The effect of meson masses is mostly additive, and the hyperfine--transverse potential exchange does not seem to be altering the order.
In the main body of the paper it can be read off, for example, in Table~\ref{TABLE-Axial-S}, among others.

So finally, we proceed to providing for reference the first and second excitations of axial vectors 
in this same calculation for closed flavor, in Tables~\ref{exc1} and~\ref{exc2}.

\begin{table}
\caption{Ground state, first and second excited states for $1^{++}$ closed--flavor quarkonium mesons in GeV. \label{exc1}}
\begin{tabular}{|llll|llll|}
\hline
$m_q$ & Ground state & First excited & Second excited\\
\hline \hline
$10^{-3}$ & $1.30$ & $1.92$ & $2.45$\\ \hline
$5\times10^{-3}$&$1.23$&$1.82$ & $2.33$\\ \hline
$10^{-1}$&$1.43$& $2.04$ & $2.54$\\ \hline
$1$&$3.65$&$4.09$ & $4.46$\\ \hline
$4$ &$9.90$&$10.20$ & $10.45$\\ \hline
\end{tabular}
\end{table}

\begin{table}
\caption{Ground state, first and second excited states for $1^{+-}$ closed--flavor quarkonium mesons in GeV. \label{exc2}}
\begin{tabular}{|llll|llll|}
\hline
$m_q$ & Ground state & First excited & Second excited\\
\hline \hline
$10^{-3}$ & $1.07$ & $1.48$ & $2.4$\\ \hline
$5\times10^{-3}$&$1.23$&$1.51$ & $2.07$\\ \hline
$10^{-1}$&$1.21$& $1.89$ & $2.43$\\ \hline
$1$&$3.59$&$4.04$ & $4.42$\\ \hline
$4$ &$9.89$&$10.19$ & $10.44$\\ \hline
\end{tabular}
\end{table}

\section{Action of the chiral charge on the $L$--$S$ basis}
For completeness, and in the notation of appendix~\ref{sec:equalflavor}, let us note how the chiral charge is acting in the basis appropriate for quarkonium $m_f=m_{f'}$.
Let us choose the following family of TDA wavefunctions (the first two with $J^{PC}=1^{--}$, with $s$ and $d$ waves respectively, the third with $J^{PC}=1^{+-}$ coresponding to the spin singlet, and the fourth to the spin triplet). Only the spin/momentum part is listed.
\begin{eqnarray}
R_1 \equiv R_s^-  & = & \frac{\boldsymbol{\sigma}}{\sqrt{2}} ,\\ \nonumber
R_2 \equiv R_d^-  & = & \sqrt{\frac{3}{2}} \left( \hat{\bf k}\cdot \boldsymbol{\sigma}\  \hat{\bf k} - \frac{\boldsymbol{\sigma}}{3} \right) , \\
R_3 \equiv R_1^+  & = & \sqrt{\frac{3}{2}} \hat{\bf k}, \\
R_4 \equiv R_3^+  & = & \sqrt{\frac{3}{2}} \boldsymbol{\sigma} \times \hat{\bf k} \ .
\end{eqnarray} 

The chiral charge acts on these wavefunctions as a multiplicative $\hat{\bf k}\cdot \boldsymbol{\sigma}$ matrix, yielding
\begin{eqnarray}
Q_5 R_1 & = & \frac{i}{\sqrt{3}} (R_3+iR_4) ,\\
Q_5 R_2 & = & \frac{1}{3} (2R_3-iR_4) ,\\
Q_5 R_3 & = & R_2 -\frac{i}{\sqrt{3}} R_4 ,\\
Q_5 R_4 & = & iR_2 - \frac{2i}{\sqrt{3}} R_1 \ .
\end{eqnarray}
It is easy to see by substitution between them that repeating the application of $Q_5$ returns the original wavefunction.

Thus, it is clear that the chiral charge is not diagonal in the $L$--$S$ basis, unlike in the mixed $s_Q$--$j_q$ useful for heavy--light systems.


\end{document}